\documentclass[useAMS,usenatbib,usegraphicx,graphics,graphicx,trackchanges,psfig]{mn2e}
\usepackage{times,color}
\input{psfig.sty}
\def\HI{\mbox{H~{\sc i}}}

\newcommand{\gimic}{{\sc GIMIC}}
\newcommand{\lya}{\hbox{Lyman-$\alpha$}}
\newcommand{\gtsima}{\hbox{ $\; \buildrel > \over \sim \;$}}

\def\Sec#1{Section~\ref{s:#1}}
\begin{document}

\title[\lya\ transmission PDF]{Sample variance and Lyman-$\alpha$ forest transmission statistics}

\author[Rollinde et al.]{E. Rollinde$^{1}$, T. Theuns$^{2,3}$,  J. Schaye$^4$, I. P\^aris$^{1}$, P. Petitjean$^{1}$\\
$^1$ UPMC Universit\'e Paris 06, UMR7095, Institut d'Astrophysique de Paris, F-75014, Paris, France\\
$^2$Institute of Computational Cosmology, Department of Physics,
University of Durham,\\ Science Laboratories, South Road, Durham DH13LE \\
$^3$ Universiteit Antwerpen, Campus Groenenborger, Groenenborgerlaan
171, B-2020 Antwerpen, Belgium\\ 
$^4$ Leiden Observatory, Leiden University, P.O. Box 9513, 2300 RA Leiden, The Netherlands\\
}

\maketitle
\begin{abstract}
We compare the observed  probability distribution function of the transmission in the \HI\ \lya\ forest, measured from the UVES \lq Large Programme\rq\  sample at redshifts $z=[2,2.5,3]$, to results from the \gimic\ cosmological simulations. Our measured values for the mean transmission and its PDF are in good agreement with published results. 
Errors on statistics measured from high-resolution data
are typically estimated using bootstrap or jack-knife resampling
techniques
after splitting the spectra into chunks. We demonstrate that these
 methods
tend to underestimate the sample variance unless the chunk size is much
larger than is commonly the case.
We therefore estimate the sample variance from the simulations. We conclude that  observed and simulated transmission statistics are in good agreement, in particular, we do not require the temperature-density relation to be \lq inverted\rq.
\end{abstract}


\section{Introduction}
At high redshift,  the intergalactic medium (IGM) contains the majority of baryons in the
Universe \citep[]{petitjeanal93,Fukugita}, is highly ionised by the UV-background (UVB)
produced by galaxies and QSOs \citep[]{GunnandPeterson} at least since redshift $z\sim 6$ \citep[]{Fan06, Becker07},  becoming increasingly neutral near $z\sim 7$ \citep{mortlock11}. It is detected in absorption against bright sources as the \HI\ \lya\ forest \citep[]{Lynds}; see \cite{Rauch98} for a review.
 
High signal-to-noise observations with high-resolution, echelle spectrographs such as the Ultraviolet and Visual Echelle Spectrograph ({\sc UVES}) on the {\em Very Large Telescope}  \citep[VLT, e.g.][]{Bergeron04} and {\sc HIRES} on Keck \citep[e.g.][]{Hu95},  of this forest of H{\sc I} absorption lines, together with numerical simulations \citep[]{Cen94, petitjeanal95, Hernquist96, Zhang95, Theuns98} and theoretical models \cite[]{Bi92,Schaye2001} have painted a picture in which low column-density H{\sc I} absorption lines trace the filaments of the \lq cosmic web\rq, and high column-density absorption lines trace the surroundings of galaxies. Simulations that include self-shielding of the UVB reproduce the observed column density distribution  over~10 orders of magnitude \citep{Altay2011}.

In this paradigm, the IGM as probed by the \lya\ forest consists of mildly non-linear gas density fluctuations. The gas traces the dark matter, and is photo-ionised and photo-heated by the UV-background. Although metals are detected in the IGM 
\citep[]{Cowie95}, even at low densities \citep[e.g.][]{Schaye03,aracilal04}, stirring of the IGM due to feedback from galaxies or AGN is probably not strongly affecting the vast majority of the baryons \citep[e.g.][]{Theuns02b,McDonald05}. 
This makes it possible to use \lya\ observations to constrain cosmological  parameters \citep[]{McDonald99,Rollinde03,Viel06,McDonald06}, as well as to probe the density distribution around quasars and galaxies \citep[]{Rollinde,Guimaraes,KimCroft}.

Photo-heating of the low-density IGM introduces a near-power law relation between its temperature, $T$, and density, $\rho$, of the form
$T=T_0\,\Delta^{\gamma-1}$, where $\Delta\equiv \rho/\langle\rho\rangle$ \citep[]{HuiandGnedin, Theuns98}. The
evolution of $T_0$ and $\gamma$ have been measured \citep{Schaye00, Ricotti2000, McDonald01,Lidzal06,Becker07,Lidzal10,Beckeral11}, and depends on the
re-ionization history \citep[e.g.][]{Theuns02a,Hui03} and the hardness of the UV-background. When the gas is strongly photo-heated after the re-ionization  of HI and HeII, $T_0$ increases and the gas becomes nearly isothermal,
$\gamma\rightarrow 1$; asymptotically the balance between photo-heating and adiabatic cooling results in $T=T_0\,\Delta^{1/1.7}$ and a slowly decreasing $T_0$ with redshift \citep[]{HuiandGnedin, Theuns98}. The amplitude of the optically thin ionising background rate ($\Gamma_{12}$), the temperature of the IGM (characterised by $T_0$ and $\gamma$), and the amplitude of fluctuations ($\sigma_8$) together determine the net amount of absorption \citep[]{Rauch97, Theuns02a, Hui03, Bolton05,Fan06,FG2008}, and the value inferred by comparing to simulations is very close to that computed by summing over sources by \cite{HM01}. 

It is also possible to compare the full probability distribution function of the transmission (TPDF) between simulations and data, which could provide a more accurate characterisation of the UVB. Such an analysis was performed by \cite{Bolton08} and \cite{Viel09}, who compared TPDFs computed from simulations to those measured from a large sample of high-resolution UVES spectra \citep[]{Kim07}. They performed a standard $\chi^2$ analysis and suggested that an \lq inverted\rq\ $T-\rho$ relation, $\gamma<1$, may be required to fit the data. A similar conclusion was reached by \cite{Becker07} using Keck data and different theoretical optical depth distributions.  \cite{Caluraal12} have done the same analysis
with additional quasars at $z\gtsima 3$. Their new analysis favours a value of $\gamma$ that is larger than what they found before, but is still slightly lower than one.
From a theory point of view it is difficult to understand how an inverted temperature-density relation might arise: simulations that include spectral hardening computed with a full radiative transfer calculation  \citep[e.g.][]{McQuinn09,Boltonal09} do not result in $\gamma<1$. If the IGM's $T-\rho$ relation were indeed inverted, 
there may be missing physics in simulations of the \lya\ forest  \citep[such as the impact of blazars as studied recently by ][]{chang2011,Puchweinal11}, which may impact  other statistics such as  the Lyman-$\alpha$ power spectrum
\citep[e.g.][]{McQuinnal11} and cosmological constraints derived from that \citep[e.g.][]{grattonal08,boyarskyal09}. Partly for this reason, Lyman-$\alpha$ forest constraints were not used by \cite{Komatsual09} in their determination of cosmological parameters from WMAP and other data. 

However, there are both numerical and observational difficulties in the characterisation of the absorption. 
Numerical issues were investigated in a paper by \cite{Tytler2009}, who analysed the importance of large-scale modes in the determination of the TPDF in a numerical simulation. These authors showed that smaller simulation boxes predict, on average, more absorption  for a given value of the imposed ionising background. The  box size used in the analyses of \cite{Bolton08}  is 56~Mpc, which, according to \cite{Tytler2009}  (their Table 12), decreases the amplitude of the TPDF by 1 to 5  per cent in the flux range used in the analysis (0.2 to 0.8) as compared to a bigger box of 76.8~Mpc. The difference could be up to 10 per cent  for even larger simulations. Even so, \cite{Tytler2009} also found that the predicted TPDF (with their box size of 76.8~Mpc) differs from the observed one, although to a lesser extent than that seen by \cite{Bolton08}.
 They did not consider an inverted temperature-density relation, but discussed other plausible sources for the discrepancy: the lack of high column density lines (log$_{10}$ $N_{\rm HI}$(cm$^{-2}$)$>$14) in the simulation,  unidentified metal lines, and the assumed mean flux values. Note that the last two  issues were discussed and, at least partly, accounted for 
 in \cite{Bolton08}.

However, an additional limitation, not considered  in \cite{Tytler2009}, is the relatively small number of observed high-resolution spectra. For example, \cite{Kim07} use a sample of just 18 spectra. In this paper we use both simulations and data to get a better handle on just how well such a relatively small sample of spectra determines the TPDF.

We revisit the analysis of the transmission statistics in terms of its
sample variance using four different observational determinations
described below in Section~\ref{s:observations}: ({\em i}) the {\sc 
  LUQAS} sample of \cite{Kim07} used by \cite{Bolton08}, ({\em ii}) the sample of \cite{Caluraal12} that increases the number of quasar with $z\gtsima 3$,
 ({\em iii}) a
sample of Keck spectra analysed and published by \cite{McDonald00}, and
finally ({\em iv}) a UVES sample collected in the context of the ESO
Large Programme \lq Cosmic Evolution of the IGM\rq\ (Bergeron et
al. 2004). 
We demonstrate that published errors on the mean transmission are often too 
small, they do not fully account for sample variance.
 The observed TPDFs are compared to mock spectra computed
from a  suite of hydrodynamical simulations called \gimic\ \citep[
Section~\ref{s:simulation}]{Crain09} that resolves both large and small
scales by using \lq zoomed\rq\ initial conditions. We generate many
mock samples from \gimic\ with the same redshift path as the observed
samples, and use this to investigate sample variance in both the mean
transmission and the transmission probability distribution. In
particular, we show how strong lines, which are relatively rare,
nevertheless have substantial impact on both the mean transmission and
its probability distribution, something which \cite{Viel04} commented
on in the context of the transmission power spectrum. 
Given the small
redshift paths of the data, we conclude that observations and
simulations are mutually consistent, because of the relatively \lq
large sample variance\rq.

\section{ Observed and simulated Lyman-$\alpha$ spectra}
\subsection{Observed samples}
\label{s:observations}
The transmission in the \lya\ forest is the ratio $F=F_{\rm o}/C$ of
the measured flux ($F_{\rm o}$) over what the flux would be in the
absence of absorption. Measuring $F$ requires knowledge of the
intrinsic flux of the quasar ($C$; the \lq continuum\rq), and since we
are only interested in absorption due to neutral hydrogen (H{\sc I}
\lya, $n=1\rightarrow 2$, $\lambda_0=1215.57~\AA$), we also need to
know the contribution to the absorption from other elements (\lq
metals\rq). Neither the continuum nor the contribution from metals are
easy to determine: the intrinsic QSO spectrum contains broad emission
lines and, moreover, the combination of a narrow slit with an echelle
spectrograph -- required to obtain the high spectral resolution --
means the spectra cannot be accurately flux calibrated.  \lq Continuum
fitting\rq\ spectra to determine $C$ then involves drawing a smooth
curve connecting regions deemed free from absorption, a somewhat
subjective procedure.  Metal lines are eliminated by identifying lines
too narrow to be due to hydrogen, or from line coincidences where a
metal transition occurs at the same redshift as a (strong) H{\sc I}
absorber or other metal transition. Finally, a \lq proximity region\rq,
{\em i.e.} the region close to the quasar where it dominates the
UV-background, is excised.

Here we use four observational data sets to determine the mean transmission and its PDF, referred to below as the LP sample, the {\em LUQAS} sample, the sample of \cite{Caluraal12}, and the M00 sample.

\begin{itemize}
\item The LP sample is from our own independent analysis of a set of 18
  {\sc UVES} VLT spectra, collected as part of the European Southern
  Observatory's \lq Large Programme\rq\ (LP) \lq Cosmic Evolution of
  the intergalactic medium\rq\ \citep[]{Bergeron04}. These LP spectra
  have a high-resolution ($\lambda/\Delta\lambda \sim$ 45000) and a high
  signal-to-noise ratio (S/N$\approx$ 25 -- 30 per pixel), and were re-binned
  on to $0.05~\AA$ pixels. The continuum was  fitted using an
  automatic method described in \cite{aracilal04}, and metal lines were
  removed by eliminating contaminated regions.  There are no damped
  \lya\ absorbers in these lines of sight. We compute the TPDFs and
  the mean transmission over three relatively small redshift ranges,
  centred at $z\simeq2$ ($1.88<z<2.37$), $z\simeq 2.5$ ($2.37<z<2.71$)
  and $z\simeq 3$ ($2.71<z<3.21$).  The total number of data pixels in
  the LP spectra for each of the redshift bins is 139830, 65067 and
  30800 (of which a fraction 74\%, 85\%\ and 100\%\ are in common with
  the LUQAS sample described below).  The corresponding absorption
  distances\footnote{The absorption distance $dX/dz \equiv (1+z)^2\,
    (\Omega_m\,(1+z)^3+\Omega_\Lambda)^{-1/2}$, and quoted numerical
    values of $dX$ assume $\Omega_m=0.25$ and $\Omega_\Lambda=0.75$.}
  are $\Delta X=10.5$, 5.8 and 2.9 respectively.

\item The {\sc LUQAS} sample used by \cite{Bolton08} and \cite{Viel09}
  is described in detail by \cite{Kim07}, including details of their
  method of continuum fitting and metal line identification.  They fit
  metal lines in the \lya\ part of the spectrum using {\sc VPFIT}
  \citep{Carswell87}, then use this to reconstruct an H{\sc I} spectrum
  without the identified metals, as in \cite{Theuns2002c}.  We find
  that this method has a similar effect on the transmission
  distribution as the method we used.  The {\sc LUQAS} sample has 18
  spectra, 14 of which are part of the LP sample.  Pixels within the
  \lya\ forest within a given redshift range are extracted and combined
  into a histogram. We will refer to these published values as the \lq
  LUQAS\rq\ data.  The transmission PDFs of \cite{Kim07} are averaged
  over the same redshift ranges as the LP ones.

\item  The \cite{Caluraal12} sample is used to investigate the TPDF  at redshift $z\gtsima 3$. Their results are  split in two bins, 
  $2.62<z<3.17$ and $3.17<z<3.72$. 
We consider the first bin only to be compared to the other determinations. 
The absorption distance in this bin, after removal of fourteen DLA and LLS regions, is about 4.5. 
We use their estimate of the  TPDF without metals and LLS.

\item The M00 sample is a set of 8 Keck {\sc HIRES} spectra with
  resolution and signal-to-noise similar to the UVES data, and is
  described in \cite{McDonald00}, hereafter M00. They use slightly
  different redshift bins that do not cover our lowest redshift bin,
  and go up to $z=4.43$. We will therefore only consider their two
  lower redshift bins: $2.09<z<2.67$ (33791 data pixels, $\Delta X\simeq 3.5$) and
  $2.67<z<3.39$ (31897 data pixels, $\Delta X\simeq 3.7$).

\end{itemize}

Noise and errors in the continuum fitting can make the transmission
$F<0$ or $F>1$.  To compute the PDF of the transmission for the LP sample,
 we use the
same binning as used in the LUQAS and McDonald et al. (2000) analyses,
{\em i.e.} bins of width 0.05 between $F=0.025$ and $F=0.975$, plus
extra bins for those pixels with $F<0.025$ and $F>0.975$. The PDF is
then normalised\footnote{Pixels with $F<0$ or $F>1$ are assigned to the
  first and last PDF bins respectively, but the number of values in
  each bin is divided by the same $\Delta F=0.05$ bin width when
  normalising the histogram.}  such that the sum of all values in all
bins equals 20. The full covariance matrix of errors on the PDF is
estimated using the jack-knife technique described in \cite{Lidzal06},
but applied to the flux, while they applied this technique to
$\delta_f\equiv (F-\bar F)/F$.  Specifically, we estimate the PDF
$P(F_i)$ from the full data sample, divide the data set into 30
different subgroups, then estimate the PDF of the data sample omitting
each subgroup iteratively, $P_k(i)$. The variance $\sigma_{i,j}$ is
then computed on the difference between $P(F_i)$ and $P_k(F_j)$:
$\sigma_{i,j}^2=\sum\limits_{k=1}^{k=30}[P(F_i)-P_k(F_i)][P(F_j)-P_k(F_j)]$.
For the other observations we use error bars taken from the
corresponding references. We discuss below
how errors can be more reliably estimated as  the variance
 among mock \gimic\ samples.  Both
estimates of errors are shown in  Fig.~\ref{fig:PDFs}, 
while Table~\ref{t:LP_pdf} indicates 
the variance among mock \gimic\ samples.

\begin{table}
\caption{The mean transmission PDF of 18 UVES Large Program (LP) QSOs,
in three redshift bins ($1.88<z<2.37$, $2.37<z<2.71$ and $2.71<z<3.21$).
The error is the 2$\sigma$ variance among mock \gimic\ samples with
ensemble average mean transmission 
$\langle F \rangle$=0.86, 0.77 and 0.71, respectively. }
\begin{center}
\begin{tabular}{cccc}
\hline
$F$  & & PDF and its error &    \\
bin centre &  $\langle z \rangle=2.0$     &  $\langle z
\rangle=2.5$      &  $\langle z \rangle =3.0$  \\
\hline
 0.00 & 0.6052 $\pm$ 0.0990 & 1.2092 $\pm$ 0.1840 & 1.6649 $\pm$ 0.4680 \\
 0.05 & 0.2004 $\pm$ 0.0390 & 0.4044 $\pm$ 0.0670 & 0.4466 $\pm$ 0.1520 \\
 0.10 & 0.1472 $\pm$ 0.0240 & 0.2734 $\pm$ 0.0390 & 0.3130 $\pm$ 0.0850 \\
 0.15 & 0.1471 $\pm$ 0.0220 & 0.2211 $\pm$ 0.0300 & 0.2894 $\pm$ 0.0700 \\
 0.20 & 0.1380 $\pm$ 0.0220 & 0.1823 $\pm$ 0.0320 & 0.2441 $\pm$ 0.0690 \\
 0.25 & 0.1370 $\pm$ 0.0210 & 0.2253 $\pm$ 0.0290 & 0.2690 $\pm$ 0.0680 \\
 0.30 & 0.1383 $\pm$ 0.0220 & 0.2228 $\pm$ 0.0300 & 0.2468 $\pm$ 0.0660 \\
 0.35 & 0.1350 $\pm$ 0.0230 & 0.2062 $\pm$ 0.0310 & 0.2527 $\pm$ 0.0620 \\
 0.40 & 0.1539 $\pm$ 0.0240 & 0.2291 $\pm$ 0.0310 & 0.2423 $\pm$ 0.0660 \\
 0.45 & 0.1602 $\pm$ 0.0260 & 0.2797 $\pm$ 0.0350 & 0.2568 $\pm$ 0.0670 \\
 0.50 & 0.1815 $\pm$ 0.0270 & 0.2780 $\pm$ 0.0340 & 0.2745 $\pm$ 0.0670 \\
 0.55 & 0.2029 $\pm$ 0.0280 & 0.2877 $\pm$ 0.0340 & 0.3474 $\pm$ 0.0750 \\
 0.60 & 0.2253 $\pm$ 0.0290 & 0.3514 $\pm$ 0.0360 & 0.4180 $\pm$ 0.0810 \\
 0.65 & 0.2855 $\pm$ 0.0320 & 0.3899 $\pm$ 0.0380 & 0.5014 $\pm$ 0.0930 \\
 0.70 & 0.3341 $\pm$ 0.0370 & 0.4519 $\pm$ 0.0420 & 0.5879 $\pm$ 0.0980 \\
 0.75 & 0.4120 $\pm$ 0.0410 & 0.5815 $\pm$ 0.0520 & 0.7192 $\pm$ 0.1210 \\
 0.80 & 0.5508 $\pm$ 0.0480 & 0.8224 $\pm$ 0.0650 & 0.8886 $\pm$ 0.1390 \\
 0.85 & 0.8279 $\pm$ 0.0610 & 1.1295 $\pm$ 0.0850 & 1.3261 $\pm$ 0.1840 \\
 0.90 & 1.3857 $\pm$ 0.0810 & 1.7072 $\pm$ 0.1070 & 1.8837 $\pm$ 0.2460 \\
 0.95 & 3.5231 $\pm$ 0.1240 & 3.1205 $\pm$ 0.1760 & 2.9413 $\pm$ 0.4360 \\
 1.00 & 10.1090 $\pm$ 0.4230 & 7.4264 $\pm$ 0.4510 & 5.8861 $\pm$ 0.8010 \\
\hline
\end{tabular}
\label{t:LP_pdf}
\end{center}
\end{table}

\subsection{Inconsistency between measured values of the mean transmission}
\label{sect:consistency}
We compare  
estimates of the mean transmission collected from the literature
\citep{McDonald00,Kirkman2005,Kim07,FG2008}, as
well as measured by us for the LP sample.
 Errors are based on a bootstrap procedure, by
resampling chunks of spectra of size 5\AA, or on the variance among
chunks of the same size \citep[][hereafter FG]{FG2008}.  \cite{Kim07}
only provide errors on the effective optical depth, for a smaller bin
in redshift $dz=0.2$. We quote the corresponding errors on the flux
$\sigma_F=F\, \sigma_\tau$, and we compute bootstrap errors for the LP
using the same bins in redshift. Estimates from LP and LUQAS are given in
Table~\ref{tab:meanf} (upper rows), with corresponding 2~$\sigma$
errors, scaled to the same absorption distance.

The mean transmission values obtained from the LUQAS and LP samples differ
  by $ 2.13,\ 2.40$ and $2.75\,\sigma$ at $z=2,\ 2.5$ and 3, respectively 
(where $\sigma$ is obtained from adding the bootstrap errors from both samples in quadrature). We
recall that the LUQAS and LP samples are mostly based on the {\em same}
raw data, but that those data were reduced by different groups. These
differences must therefore be due to systematic errors in the adopted
procedures, in particular differences in continuum fitting and the
treatment of absorption from metals. Also, \cite{Kim07} concluded that
the treatment of the data, in particular continuum fitting, leads to
notable differences between authors. Published values for $\bar F$ from
LUQAS, \cite{Kirkman2005} and FG agree within 1~$\sigma$ at $z=2$, but
the differences increase at higher $z$. The most discrepant values are
2.49$\sigma$ at $z=2.5$ \citep[LUQAS versus][ both are
high-resolution data]{McDonald00}, and 3.9$\sigma$ at z=3 \citep[][\, versus FG]{Kirkman2005}.

 How reliable are the quoted errors? \cite{Kim07} estimate errors
  on the effective optical depth, $-\ln(\bar F)$, by bootstrapping the
  LUQAS spectra in chunks of 5\AA. They do not mention convergence
  tests with chunk size for the error on the mean flux, but they do note that a
  modified jack-knife method, using 50~\AA\ chunks, yields errors that
  are too low -- comparable to the estimated variance due to continuum
  placement alone. They nevertheless use jack-knife errors with 50~\AA\
  chunks to compute the variance of the transmission PDF.  
 \cite{Caluraal12} compare errors  on the TPDF estimated with a
 bootstrap on 5~\AA\  chunks and with a  jack-knife on 50~\AA\ chunks. 
They find similar results, but do not mention convergence
  tests with chunk size either. 
  FG (2008)
  mention that "We have verified that the error estimates have
  converged for our choice of segment length", but they do not present
  quantitative results. 

 Bootstrap errors  depend on the arbitrary size of the chunks
  from which they are computed.  Indeed, for the LP data at
  $z=2.5$, we find variances in the mean flux of
  $\sigma=[0.25,0.53,0.78,1.14, 1.03,1.33,1.15,1.16]\times 10^{-2}$ for chunk sizes
  of $[0.2, 1, 5, 25, 50, 125, 250, 625]$~\AA. Although $\sigma$ converges for
  very large chunk sizes $\gtsima 25$~\AA, as expected, we suggest that
  typical published errors based on 5~\AA\ chunks underestimate the
  variance by $\sim 50$~per cent. Note that the largest chunk size we
  tested, 625~\AA, is comparable to the extent of the \lya\ forest in a
  $z\sim 3$ QSO. We discuss the reliability of bootstrap errors using
  \gimic\ mocks further in Section~\ref{s:mock-error} below.

\subsection{Mock samples}
\label{s:simulation}

\begin{figure}
\includegraphics[width=\linewidth]{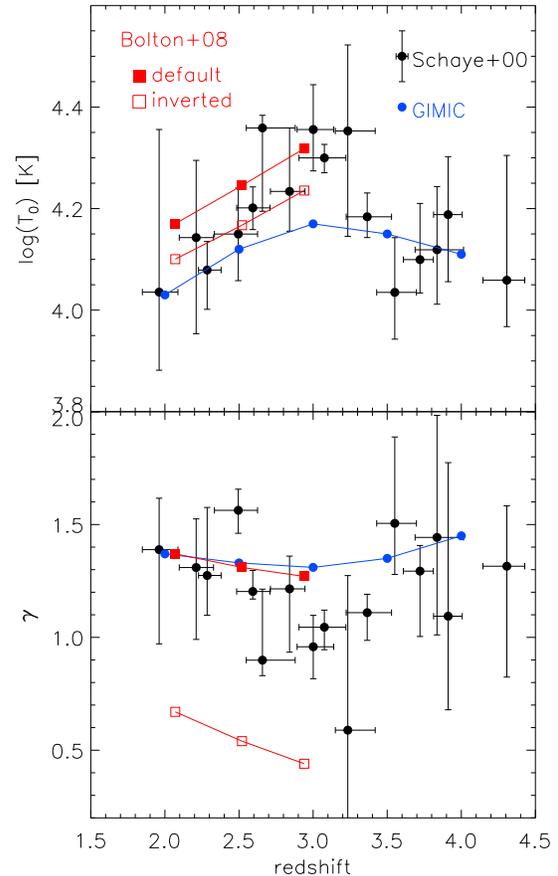}
\caption{Evolution of the parameters $T_0$ and $\gamma$ of the
  temperature-density relation
  $T=T_0\,(\rho/\langle\rho\rangle)^{\gamma-1}$, as measured by Schaye
  {\em et al.} (2000, black circles with error bars) and in the \gimic\
  simulation (blue connected dots). The temperature-density relation in
  \gimic\ is broadly consistent with the measured values. HeII
  re-ionization causes the rise in $T_0$ and the corresponding dip in
  $\gamma$ in the \gimic\ simulations at redshift $z\sim 3.2$, but
  $\gamma$ never drops below $\sim 1.3$.  Red symbols are from the
  model of Bolton {\em et al.} (2008): filled squares are for their
  default model, open squares are for their model 20-256g3 that best
  fits the transmission PDF they inferred from LUQAS. This model has an
  inverted temperature-density relation, {\em i.e.} $\gamma<1$.}
\label{fig:T-evol}
\end{figure}

We use the \gimic\ \citep[Galaxies-Intergalatic Medium Interaction
Calculation,][]{Crain09} simulations, a set of smoothed particle
hydrodynamics simulations (SPH) of five nearly spherical regions of
co-moving radius $R\sim 18h^{-1}\,{\rm Mpc}$ picked from the Millennium
simulation \citep[]{Springel05}. The simulations have a gas particle
mass of $1.4\times 10^6\,h^{-1}\,{\rm M}_\odot$~. These \lq zoomed\rq\
simulations allows us to obtain high numerical resolution yet include
the effects of large-scale power, {\em i.e.} the simulation probes a
range of environments, from massive clusters to deep voids. The effect
of large-scale structures, as discussed in \cite{Tytler2009}, is thus
accounted for.

The \gimic\ simulations were performed with the {\sc Gadget-3} code, an
evolution of {\sc Gadget-2} described last by \cite{Springel05b}, with
modules for star formation, feedback from galactic winds,
chemo-dynamics, and radiative cooling and photo-heating due to an
imposed evolving UV-background, as described in
\cite{SchayeandDallaVecchia,DallaVecchiaandSchaye} and Wiersma et
al. (2009b, a), respectively, see also \cite{schayeal10}. The assumed
cosmological parameters are $(\Omega_{\rm
  cdm}+\Omega_b,\Omega_\Lambda,\Omega_b,n_s,h,\sigma_8)=(0.25,0.75,0.045,1,0.9,0.73,0.9)$.
The five \gimic\ regions are picked such that their over-densities at
redshift $z=1.5$ are $(-2,-1,0,1,2)$ times the root-mean-square
deviation, $\sigma$, from the mean on the spatial scale of the spheres.
Re-ionization of H{\sc I} is assumed to occur at $z=9$, heating the IGM
to $T\sim 10^4$~K, and of He{\sc II} at $z=3.5$. As also shown by
Wiersma et al. (2009b), the evolution of $T_0$ and $\gamma$ in the
simulations is broadly consistent with the \cite{Schaye00}
measurements, see also Fig.~\ref{fig:T-evol}. For densities close to
the mean, $\gamma\gtsima 1.3$, and the temperature-density relation is
never \lq inverted\rq.

We compute  1000 mock \lya\ forest spectra by tracing straight lines
through a cube\footnote{The cubes have sides $\sim
  11\,h^{-1}$~co-moving Mpc which ensures we stay well away from the
  edges of the spheres to avoid artificial boundary effects, see
  \cite{Crighton2010} for details.  We will call a \lya\ spectrum
  obtained from a single cut through the cube a {\em short} spectrum.}
embedded well within each of the five spheres, extracting
density, temperature and peculiar velocity along them, and then
computing the corresponding optical depth as described in
\cite{Theuns98}. \cite{Crain09} explain in their appendix how to
combine results from individual spheres to correctly reproduce
statistics valid for the full Millennium volume: we use the weights
listed in their Table A1. Given these weights, we generate a \lq
mock\rq\ LP sample by randomly selecting spectra from each of the five
spheres until the redshift path of mock and LP samples are the same.
We repeat this procedure 400 times to obtain a \lq suite\rq\ of mock
samples. Note that every single mock sample in the suite has the same
redshift path as the LP sample. Each spectrum is convolved with a
Gaussian to match the {\sc UVES} spectral resolution, re-binned to the
UVES pixel size, and we add noise with similar statistical properties
as measured in the observed spectra.  Our results do not change
  significantly if we only use the \gimic\ mean density sphere.  We can
  compute flux statistics for a given mock sample simply from all
  pixels in all short spectra that make-up the mock sample. However,
  when computing bootstrap errors below, we combine these short spectra
  into a \lya\ spectrum that mimics the full absorption distance of a
  given LP spectrum.

It is difficult to accurately mimic the effect of \lq continuum
fitting\rq\ as applied to observations to the simulated samples,
because the wavelength range over which the observed continuum is
supposed to vary is large compared to the size of an individual
simulated spectrum. In the observations, the true and estimated
continua are thought to differ by about 1--3 per cent \citep[see
e.g.][]{aracilal04,FG2008}. Therefore, to investigate plausible
continuum uncertainties, we compare statistics from the original
samples to those in which we multiply the flux by a constant factor of
1.02 to mimic a 2 per cent systematic offset between \lq true\rq\ and
\lq fitted\rq\ continua.

The \lya\ optical depth in a spectrum depends on the evolving
photo-ionization rate,
\begin{equation}
\Gamma=4\pi\int_{\nu_T}^{\infty}\frac{J(\nu)}{h \nu}\sigma_\nu\,{\rm d}\nu\  \,\equiv \Gamma_{12}\,10^{-12}\, {\rm s}^{-1}\,,
\label{eq:gamma}
\end{equation}
where $J(\nu)$ is the mean intensity of the ionising radiation at a
given redshift, $\nu_T$ is the frequency of the Lyman limit,
$\sigma_\nu$ is the hydrogen photo-ionization cross section. Within a
suite of mock samples we use the same value for $\Gamma_{12}$, and will
refer to the \lq ensemble average\rq\ mean transmission of the suite as
$\langle F\rangle$. The mean transmission, $\bar F$, of a given mock
sample can differ significantly from the ensemble average $\langle
F\rangle$ of the corresponding suite because of \lq sample variance\rq\,
 and the same is true for its PDF.  We estimate the sample variance in
a given suite by comparing all 400 mock samples that make-up the
suite. We emphasize that because the simulated samples keep probing the
same density field, the real dispersion is likely to be larger than
this estimate.

The value of the photo-ionization rate $\Gamma_{12}$ is
uncertain. \cite{Theuns98} show that in the optically thin case,
simulations can be run with one value for $\Gamma_{12}$ and later
accurately scaled to another value.  To investigate the effect of
uncertainties in $\Gamma_{12}$, we generate many suites of mock
samples, with different values of $\Gamma_{12}$ and hence of the
ensemble average transmission, $\langle F\rangle$.

\subsection{Estimates of errors with mock samples}
\label{s:mock-error}
We can check the reliability of the  bootstrap errors discussed in
Section~\ref{sect:consistency} using \gimic\ mock samples. We
  first examine whether mocks generated from the simulation give the
  same errors on the mean flux as observed samples when the errors are
  estimated in the same way.
 \cite{FG2008} divide the variance $\sigma_i$ of the mean flux
  measured along chunks of 3~Mpc proper size, by the square root of the
  number of chunks. They find $\sigma_i=[0.13, 0.11,0.09]$ at
  $z=[3,2.4,2]$, %and their redshift path corresponds to 
 with 193, 263 and 50 chunks respectively. Applying
  this procedure first to the LP data, we find $\sigma_i=[0.125,
  0.13,0.095]$ at $z=[3,2.5,2]$, %and their redshift path corresponds to
with  37, 262 and 413 chunks
  respectively. Applied to our mocks we find $\sigma_i=[0.14,
  0.14,0.11]$. Therefore both our analysis of the LP observations, and
  of the \gimic\ simulations, give error estimates in reasonable agreement
  with those obtained by \cite{FG2008}.  \cite{Kim07} estimate errors
  on the effective optical depth, $-\ln(\bar F)$, by bootstrapping the
  LUQAS spectra in chunks of size 5~\AA.  We concentrate on their
  estimate at $\bar z=2.59$ with a bin in redshift of $\Delta z=0.2$,
  corresponding to a velocity path of 88682 km~s$^{-1}$. We use the
  \gimic\ simulations to generate many mock versions of the LUQAS
  sample, each with the same velocity path, and estimate the variance
  $\sigma$ for the same chunk size. The average value for our mocks
  is $\sigma_F=F\, \sigma_\tau=0.0124$, identical to their bootstrap
  error.  Finally, we compare errors estimated from \gimic\ against
  our own bootstrap errors obtained from the LP data, as discussed in
  the previous section.  At $z=2.5$ and for a velocity path of
  $\sim$~190000~km~s$^{-1}$), we calculate bootstrap variances of
  $\sigma=[0.26,0.54,0.80,0.98,1.22,1.15]\times 10^{-2}$ for chunk
  sizes of $[0.2, 1, 5, 25, 125, 625]$~\AA\ for the simulated mocks, as
  compared to $\sigma=[0.25,0.53,0.78,1.14,1.33,1.16]\times 10^{-2}$ for
  the LP observational data. We conclude that errors computed from
  \gimic\ mocks are in excellent agreement with published errors, as
  well as errors obtained by us from the LP data, when simulated and
  observed errors are calculated in the same way.

The bootstrap errors discussed above clearly depend on the value
  of the chunk size for which they are computed, both for the data and
  for the simulated spectra. They start to converge for relatively
  large chunk sizes of $\sim$~25~\AA, although the convergence is not yet clearly reached.
 Using simulations we can also
  calculate the variance between different mock samples: simply
  generate many mock samples for a given simulation, each with the same
  redshift path as a given observed sample, and evaluate the variance
  between mock {\em samples}. This variance is $[0.55,0.88,1.7]\times
  10^{-2}$ at redshifts $z=[2,2.5,3]$, as compared to bootstrap errors
  using 25~\AA\ chunks of $[0.50,0.98,2.1]\times 10^{-2}$, in reasonable
  agreement. Given the dependence of the variance on chunk size for
  small chunks, we will use the variance between mock samples to
  characterise the expected level of scatter in the data and to
  investigate the consistency between simulation and data. 
We suggest that  error estimates that we obtain from determining the variance between mocks, are
more realistic than the published, observed bootstrap errors.

\section{The transmission PDF}

\label{s:tfpdf}

We have computed the transmission PDFs of the LP sample over the same
small redshift ranges as used by \cite{Kim07}. Because these redshift
ranges are relatively narrow, evolution over them can be safely
neglected, and hence we simply use simulation snapshots at a single
redshift \citep[$z\simeq$ 2, 2.5 and 3 for the three bins used by
][]{Kim07} when comparing to the observed data.

\begin{figure}
\includegraphics[width=\linewidth]{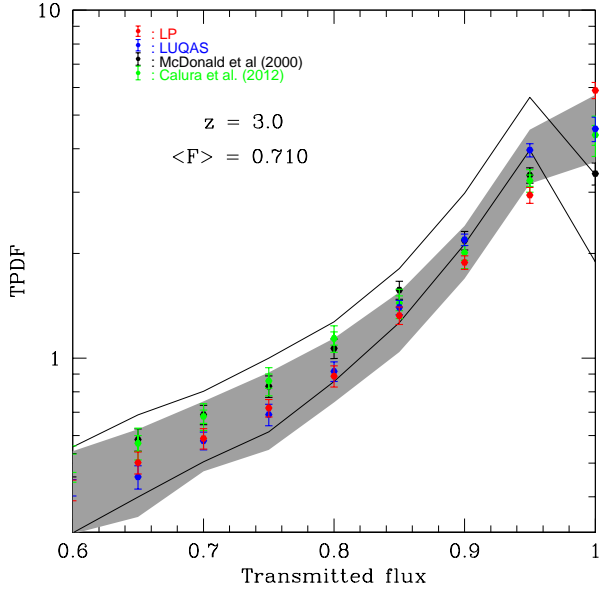}
\caption{Effect of \lq continuum fitting\rq\ the \gimic\ simulations,
  {\em solid curves} show the 2$\sigma$ range in the transmission
  PDFs of a sample of mocks with given ensemble averaged transmission,
  $\langle F\rangle$.  When errors in continuum fitting are mimicked by
  a systematic shift in the continuum (see Section 2.2), the range
  is enclosed by full lines.  Continuum fitting makes the
  shape of the TPDF uncertain close to $F= 1$. Note that we only show
  the range $0.6\le F \le 1$. For $F<0.6$, we find that the continuum
  correction is small compared to the 2$\sigma$ range.  Symbols with error
  bars are the data from the LP sample ({\em red}); LUQAS ({\em blue}),
  McDonald et al. (2000) ({\em black}) and Calura et al. (2012) ({\em green}).
 These also show significant
  differences in the range $F>0.7$, plausibly due to the different
  continuum fitting methods applied in the data reduction.}
\label{fig:PDF_cont}
\end{figure}

\begin{figure}
\unitlength=1cm
\begin{picture}(8,18.5)
\put(0,-0.4){\psfig{width=7cm,figure=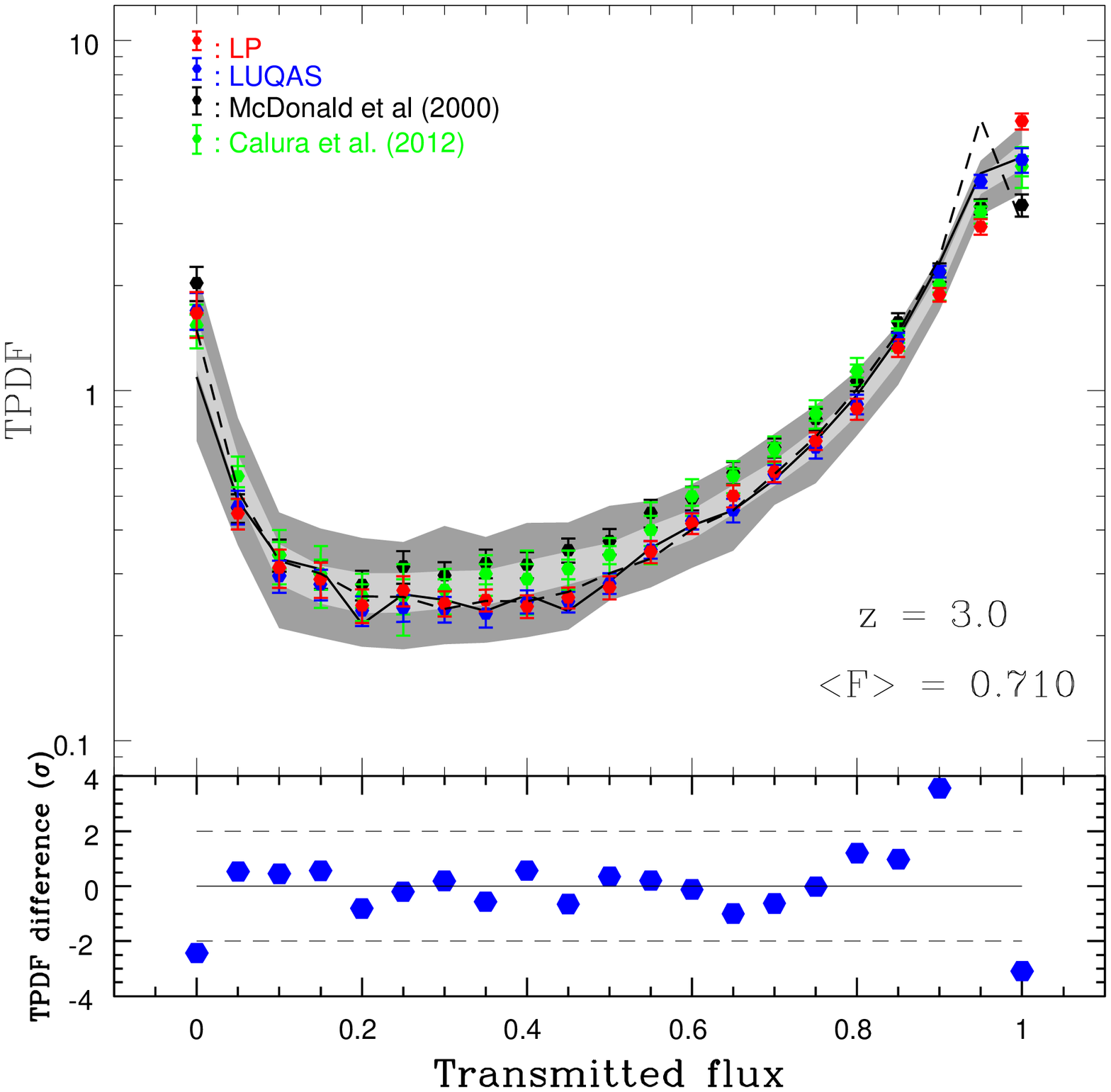}}
\put(0,6.){\psfig{width=7cm,figure=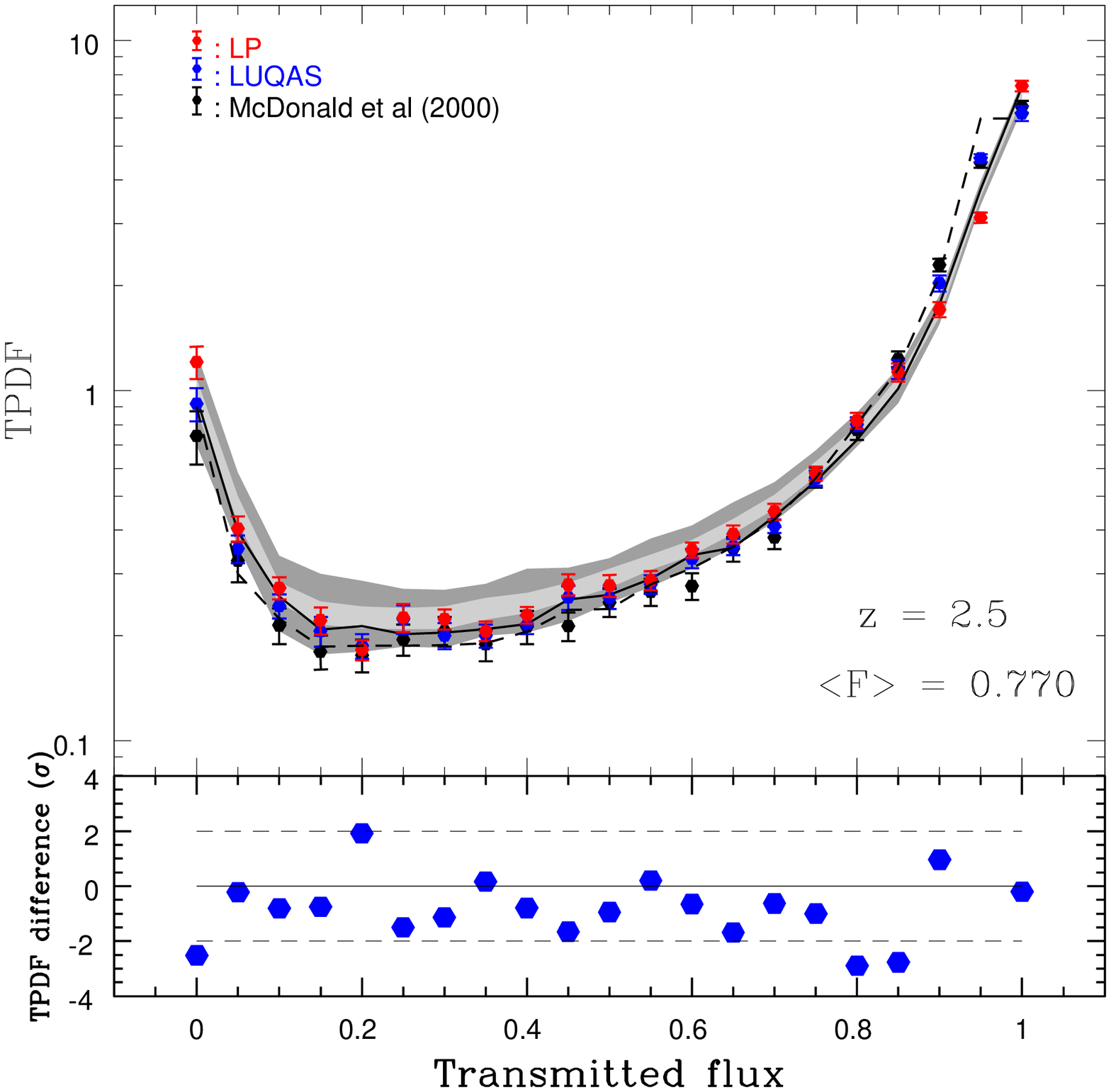}}
\put(0,12.4){\psfig{width=7cm,figure=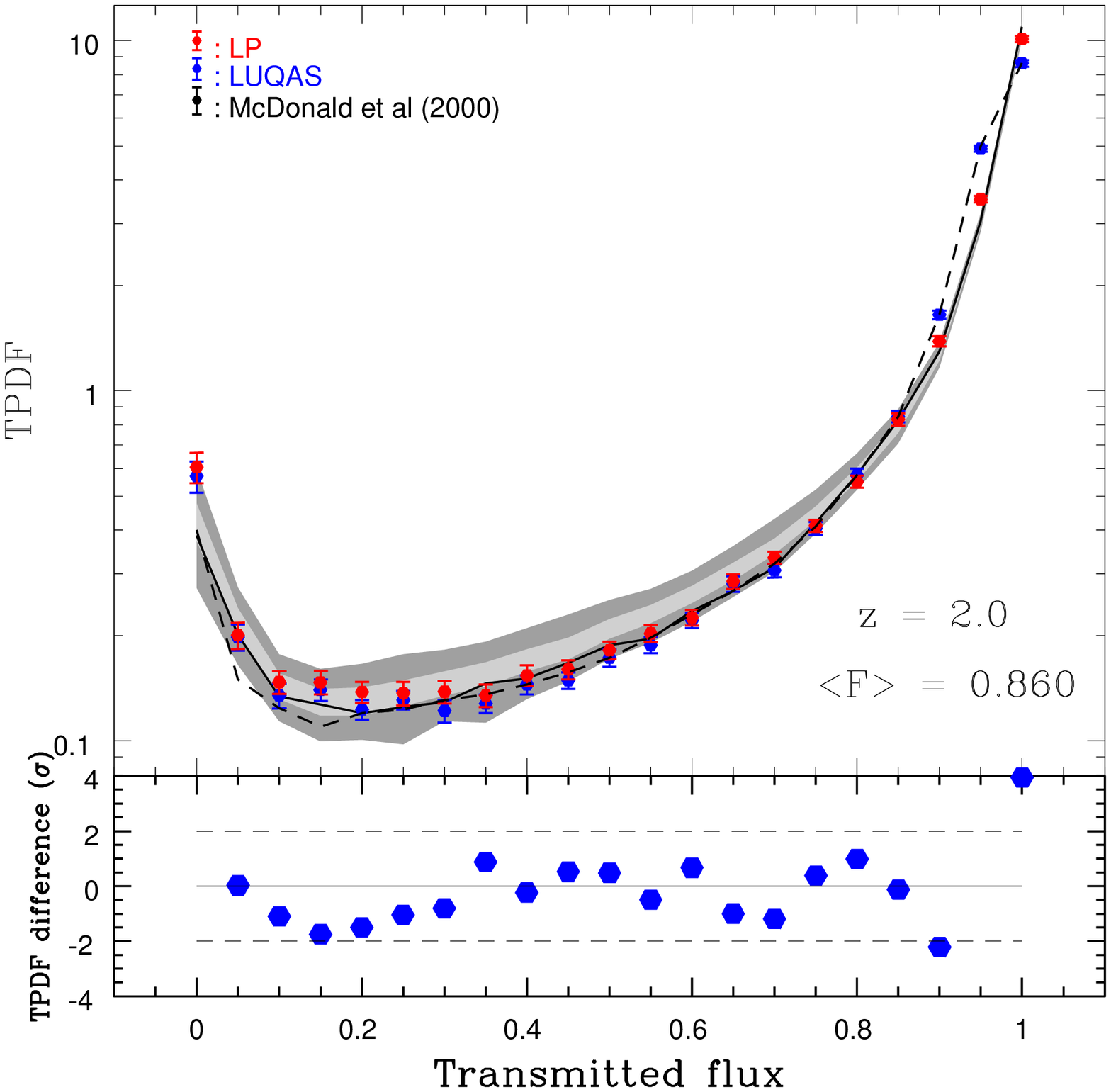}}
\end{picture}

\caption{{\em Top to bottom}: PDF of the transmission at ${\bar z}=2$,
  2.5 and 3, of the best-fitting simulations (continuum fitted \gimic\
  simulation:{\em solid curve}; Bolton et al. 2008 model 20-256g3 shown
  as open squares in Fig~1: {\em dashed curve}), compared to
  observational data (symbols with error bars, LP sample: {\em red};
  LUQAS: {\em blue}; Calura et al.: {\em green}; M00: $z$=2.41 and 3.0, {\em black}).
 Error bars are  $1\sigma$ jack-knife 
  errors for LUQAS, LP and Calura et al.,
 and bootstrap of 5\AA\ chunks for M00.
 Light (dark) shaded regions
  correspond to the 1 and 2$\sigma$ range computed from 400 mock
  LP samples in \gimic\ simulations with redshift and ensemble
  averaged mean transmission $\langle F\rangle$ as indicated in each
  panel. The simulations and various data sets agree well within the
  2$\sigma$ range at all three redshifts. Insets
  show (model-data)/$\sigma_o$, where model is the best-fitting PDF for \gimic,
 data and  $\sigma_o$ are the LP PDF and the variance estimated in \gimic\ simulations. The
  \gimic\ simulations fit the data for $F<0.7$ even though $\gamma>1$
  at all $z$.  For $F > 0.7$ and $z \leq 2.5$, different data sets
    are inconsistent and sensitive to continuum fitting (missing points are above 4).}
\label{fig:PDFs}
\end{figure}

\begin{figure}
\includegraphics[width=\linewidth]{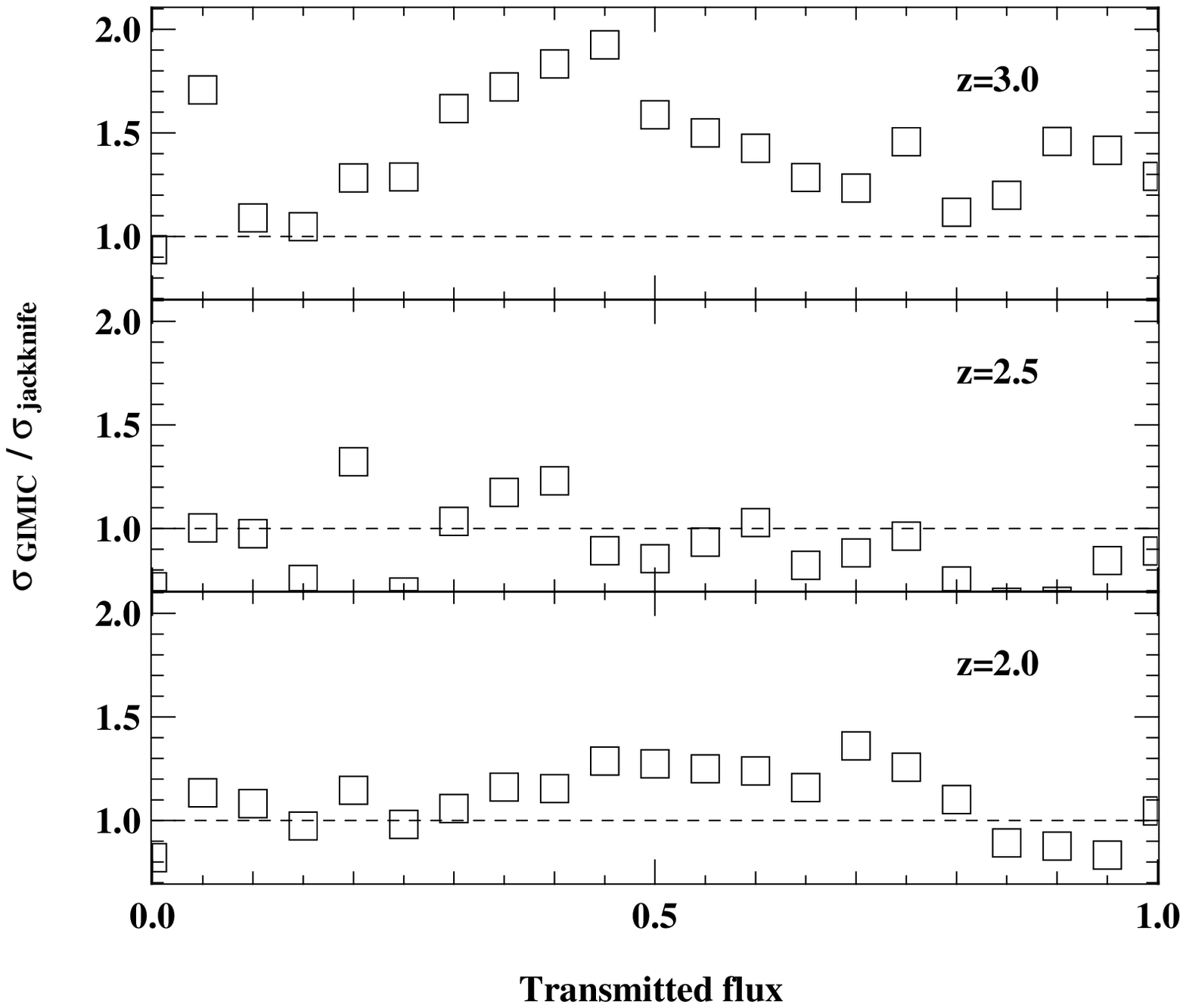}
\caption{Ratio of PDF variance  computed from 
\gimic\ mocks to variance computed from jack-knife method for bins in
 transmitted flux
 in three different bins in redshift.}
\label{fig:pdf_error}
\end{figure}

\subsection{Variance of the transmission PDF} 
Fig.~\ref{fig:PDF_cont} illustrates that continuum fitting quite
noticeably affects the transmission PDF near $F\simeq 1$, and
comparison to the over-plotted data also suggests that uncertainties in
continuum placement can explain the large differences in the {\em
  observed} PDFs at $F\simeq 1$. Recall that we mimic the errors in
continuum fitting by a systematic shift in the continuum (Section 2.2).
Clearly, given these uncertainties, this part of the TPDF cannot
constrain models robustly \citep[see also][]{Meiksin2001}. Fortunately,
the distribution of pixels with $F<0.7$, say, is relatively insensitive
to the error in the continuum placement for high-resolution spectra and
can thus be used to constrain the mean transmitted flux.

The \gimic\ simulations that best reproduce the observed transmission
PDFs for $F<0.7$ have ensemble averaged mean transmissions of $\langle
F \rangle$=0.86, 0.77 and 0.71 at redshifts z=2,\,2.5 and 3,
respectively, as discussed in more detail below. Observed and mock
TPDFs with these values of $\langle F \rangle$, are compared at
$z=2,\,2.5$ and $3$ in Fig.~\ref{fig:PDFs}. Light (dark) shaded regions
show the 1$\sigma$ and 2$\sigma$ dispersion\footnote{ They correspond
  to the 2.275, 15.8655, 84.13 and 97.725 percentiles computed from 400
  realisations.}  among TPDFs of this particular suite of mocks. There is considerable variance between the transmission PDFs of mock realisations, even though each mock realisation is generated from the same simulation with the full absorption distance of the LP observed sample.

The variance in the mocks increases with redshift since the redshift
path decreases. 
The ratio of variance computed from \gimic\ mock versus jack-knife variance 
is shown in Fig.~\ref{fig:pdf_error}. Except at $z=2.5$, variance in mocks is systematically larger, from 10 to 50\%\ at $z=2$ and up to 100\%\ at $z=3$. 
 Given that the simulations, if anything, {\em
  underestimate} sample variance, suggests once more that the
observationally determined jack-knife errors are too small. 
Although more difficult to assess from other works, we found that the estimates 
of errors using the jack-knife method is very unstable given the relatively small size of the sample.
We will therefore quote variances computed from our mocks only.

The LP and LUQAS data fall well within the 2$\sigma$ region at all $z$
for $F<0.7$, with a possible exception of the $F\simeq 0$ bin at
$z=2$. It is possible that the latter discrepancy is due to the fact
that simulations that assume the gas to be optically thin do not
reproduce the observed number of strong lines
\citep[e.g.][]{Tytler2009}. Including self-shielding appears to solve
this issue \citep{Altay2011}. The LP and LUQAS samples results 
are almost identical in bins where uncertainties in the position of
the  continuum  does not interfere in the TPDF. 
They are  also very similar to the results from \cite{Caluraal12} sample
that has one quasar in common (which makes one fourth of the total
sample in this redshift bin). They  also agree with
results from \cite{McDonald00} within the 2$\sigma$ range estimated
from the simulations.
 
The difference between the best-fitting simulated PDFs 
in \gimic\ mock samples
(among different values for $\Gamma_{12}$ only) 
and our determination of the  TPDF from the LP,
divided by $1\sigma$ range on mock LP  TPDF in  \gimic\ simulation,
 is shown in the bottom of each
panel in Fig.~\ref{fig:PDFs}. 
There is no evidence
that the observed and simulated \gimic\ PDFs are inconsistent at any
redshift.  
The statistical interpretation of this measurement, and the
derived constraints on the ionising background rate, are discussed
further in \Sec{gamma}.

\begin{figure}
\includegraphics[width=\linewidth]{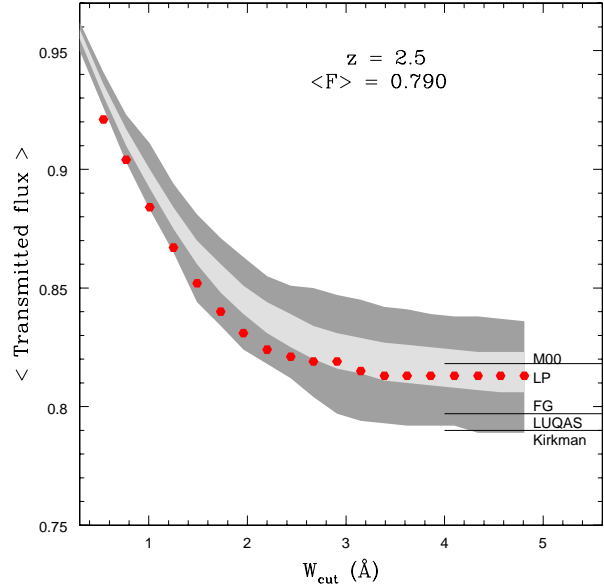}
\caption{Mean transmission of spectra that include all lines with
  equivalent width $W<W_{\rm cut}$ for the LP sample (red dots), and
  the corresponding 1 and 2$\sigma$ range in this quantity estimated
  from \gimic\ mock samples (grey and dark regions, respectively). The
  net mean transmission values, $\bar F$, for the LUQAS, M00,
  Faucher-Gigu\`ere et al. (2008; FG) and Kirkman et al. (2005) data
  are indicated by horizontal lines (FG and LUQAS values of $\bar F$
  are identical).
% with 1$\sigma$ error bars for M00, Kirkman (since
%  their sample is smaller, error bars are larger) and FG. 
There is
  significant scatter in $\bar F$ of the \gimic\ samples when $W_{\rm
    cut}\gtsima 1.5$ \AA, but as strong lines are excised, the
  dispersion decreases significantly.  This shows that strong lines are
  mostly responsible for the scatter. The {\em observed} values of the
  net mean transmission, $\bar F(W<\infty)$, are well within the
  2$\sigma$ range estimated from the \gimic\ simulations.}
\label{fig:fWs}
\end{figure}

\subsection{Variance of the mean transmission}  

Interestingly, observations as well as simulations show large
quasar-to-quasar variations in the mean transmission at a given
redshift. To illustrate the origin of this large scatter, we analyse
400 mock samples from \gimic\ generated with a given ensemble average,
$\langle F\rangle=0.79$, at redshift $z=2.5$. The large scatter is due
to strong absorption lines, which contribute significantly to the mean
opacity: the small number of strong lines per QSO spectrum introduces
the observed scatter, as we now show \citep[see
also][]{Desjacques2007}.

We have used a simple criterion to identify \lq lines\rq\ in the
spectrum as regions between two maxima in $F$; we also demand that the
corresponding minimum is sufficiently different from the lowest maximum
to avoid identifying noise features as lines.  More specifically, this
algorithm identifies all local minima and maxima on a spectrum smoothed
with a Gaussian kernel of width 8~km~s$^{-1}$. A line consists of all
pixels between two maxima that satisfy the following two conditions:
($i$) two successive maxima must be separated by more than
8~km~s$^{-1}$ and ($ii$) the flux difference between the maxima and the
minimum they straddle must be larger than four times the estimated
error per pixel. Each pixel is then assigned to a line, with given
equivalent width, $W$. We can now compute the mean transmission in a
mock sample (or the LP data) for all pixels in lines with $W$ less than
some maximum equivalent width, $W_{\rm cut}$.

The mean transmission, $\bar F(W_{\rm cut})$, for all pixels in lines
weaker than a given value of $W_{\rm cut}$ is plotted as a function of
$W_{\rm cut}$ in Fig.~\ref{fig:fWs} as red dots for the LP sample, with
grey and dark regions the 1 and 2$\sigma$ range estimated from the
mock \gimic\ samples.  For a high cut in $W$, all pixels are used and
$\bar F(W_{\rm cut}=\infty)$ is simply the net mean transmission $\bar
F$; we also indicate $\bar F$ from LUQAS, M00, FG and
\cite{Kirkman2005}.

For mock samples with ensemble average $\langle F\rangle=0.79$ we find
that the (continuum fitted) $\bar F(W_{\rm cut}=\infty)$ varies between
0.79 and 0.84 within 2$\sigma$. Note that our procedure to estimate the
errors due to \lq continuum fitting\rq\ makes the mean transmission,
$\bar F$ systematically higher than $\langle F \rangle$.  Observed
determinations of the mean transmission are shown with horizontal lines
in the figure.
%; their values are also listed in
%Table~\ref{tab:meanf}. 
It appears that, despite the large dispersion
amongst observed values, they are nevertheless consistent, because the
expected sample variance, as inferred from \gimic\ (and consistent with 
bootstrap estimates using real data for sufficiently large chunk size),
 is so large. The
origin of the large variance is the presence of strong lines. 
%When
%those are ignored, the variance in mock samples decreases considerably,
%demonstrating that it is the Poisson noise in absorption due to \lq
%strong\rq lines that causes the large scatter.

\begin{figure}
\includegraphics[width=\linewidth]{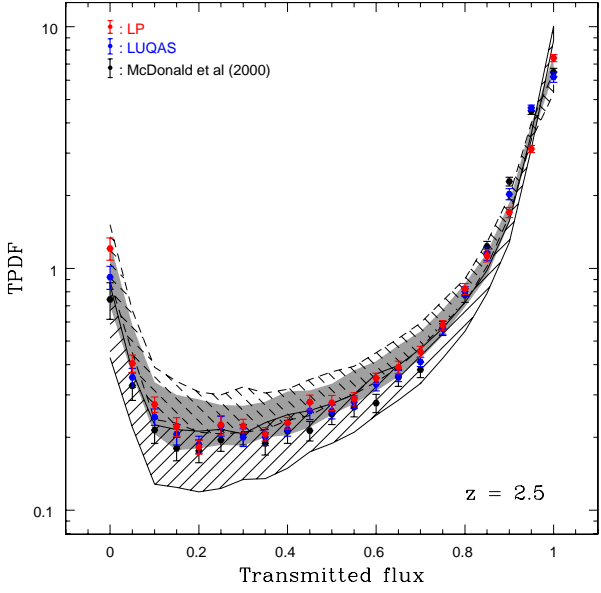}
\caption{Dependence of the transmission PDF on the ensemble averaged
  $\langle F\rangle$ at $z=2.5$.  The dark shaded region shows the
  2$\sigma$ range computed from 400 mock samples in a \gimic\
  simulation with $\langle F\rangle=0.77$ as in Fig.~\ref{fig:PDFs};
  Symbols with error bars are as in Fig.~\ref{fig:PDFs}. Solid and
  dashed hashed regions correspond to the 2$\sigma$ range in
  \gimic\ simulations with $\langle F\rangle=0.83$ and 0.74,
  respectively.  At these extremes the observational data (for
  transmission $0.1<F<0.7$) falls just outside the 2$\sigma$ range of
  the simulation for at least one data point.}
\label{fig:pdf_favg}
\end{figure}

\begin{figure}
\includegraphics[width=\columnwidth]{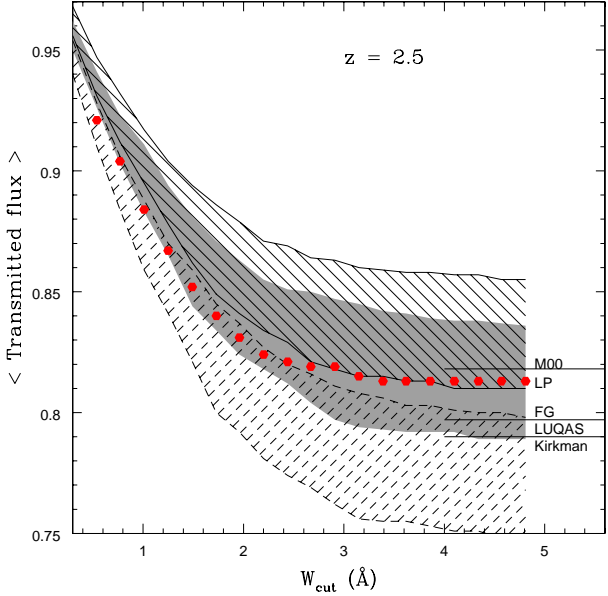}
\caption{Same as Fig.~\ref{fig:pdf_favg}, but for the dependence of the
  mean transmission as a function of maximum line width, $\bar F(W_{\rm
    cut})$.  Dark shaded region is the $2\sigma$ range for $\langle
  F\rangle=0.79$, solid and dashed hashed regions correspond to the
  2$\sigma$ range in the \gimic\ simulations with $\langle
  F\rangle=0.81$ and 0.75, respectively.  
}
\label{fig:fW_favg}
\end{figure}

\begin{figure}
\unitlength=1cm
\begin{picture}(8,18.5)
\put(0,-0.4){\psfig{width=7cm,figure=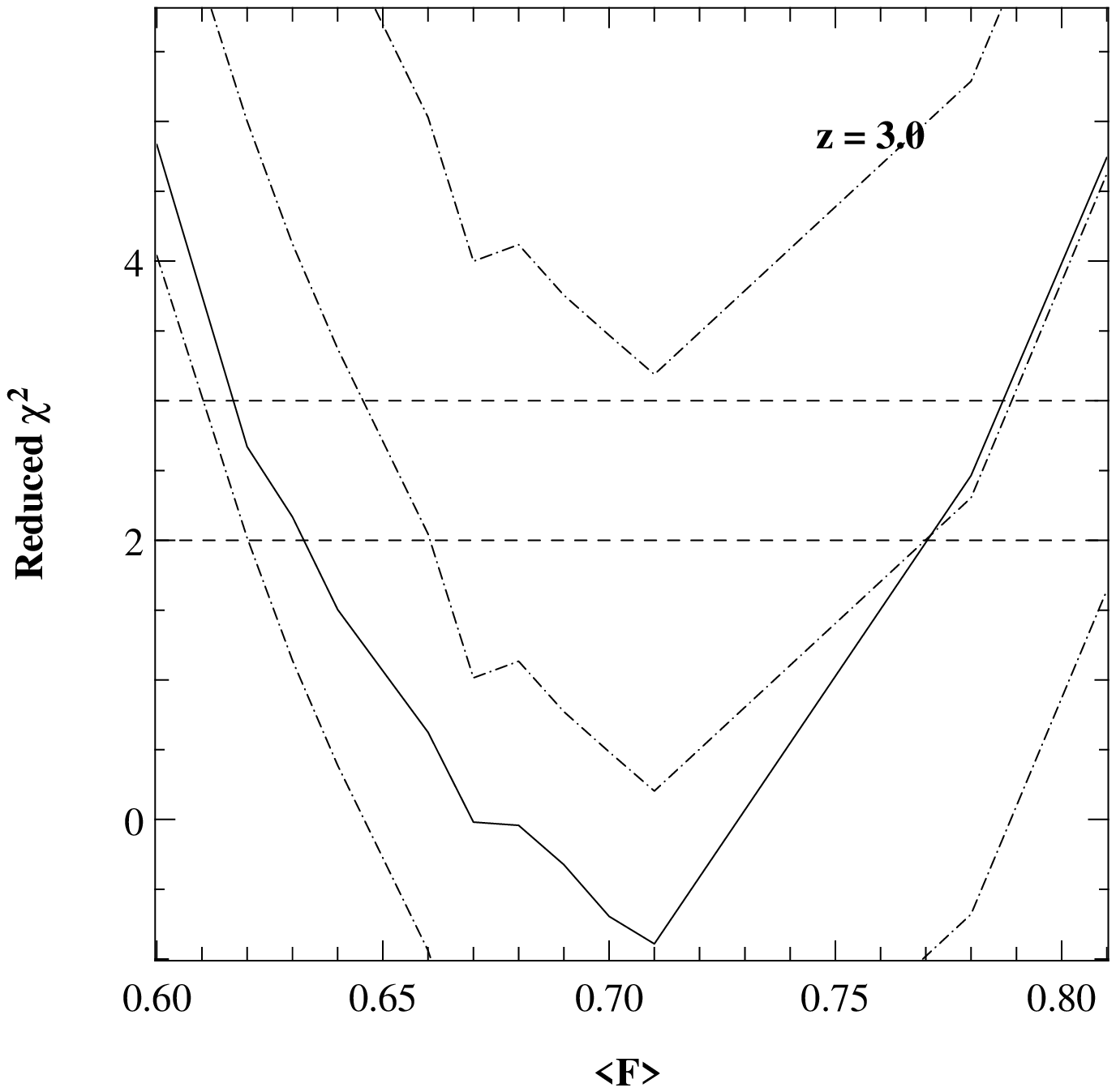}}
\put(0,6.){\psfig{width=7cm,figure=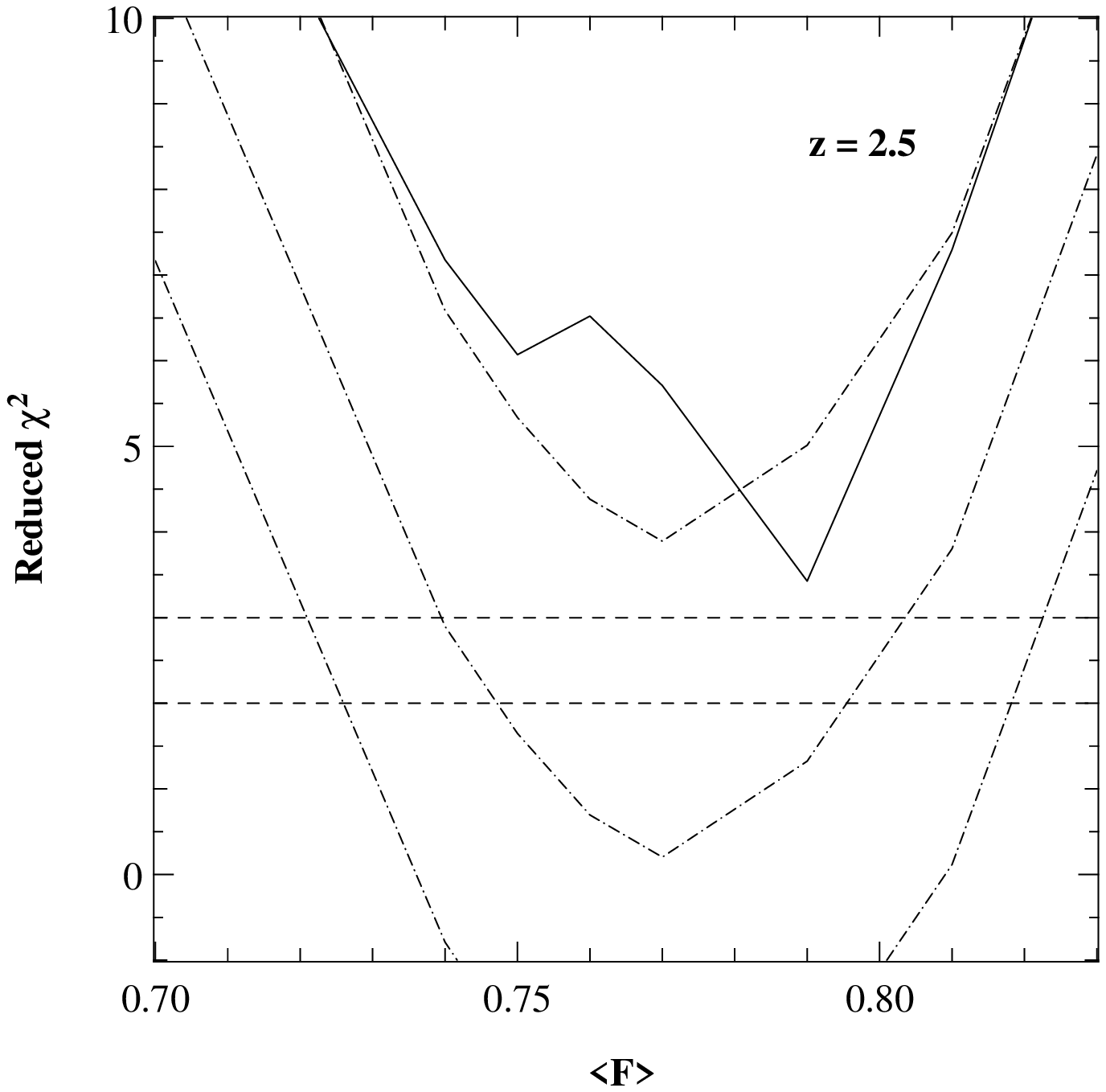}}
\put(0,12.4){\psfig{width=7cm,figure=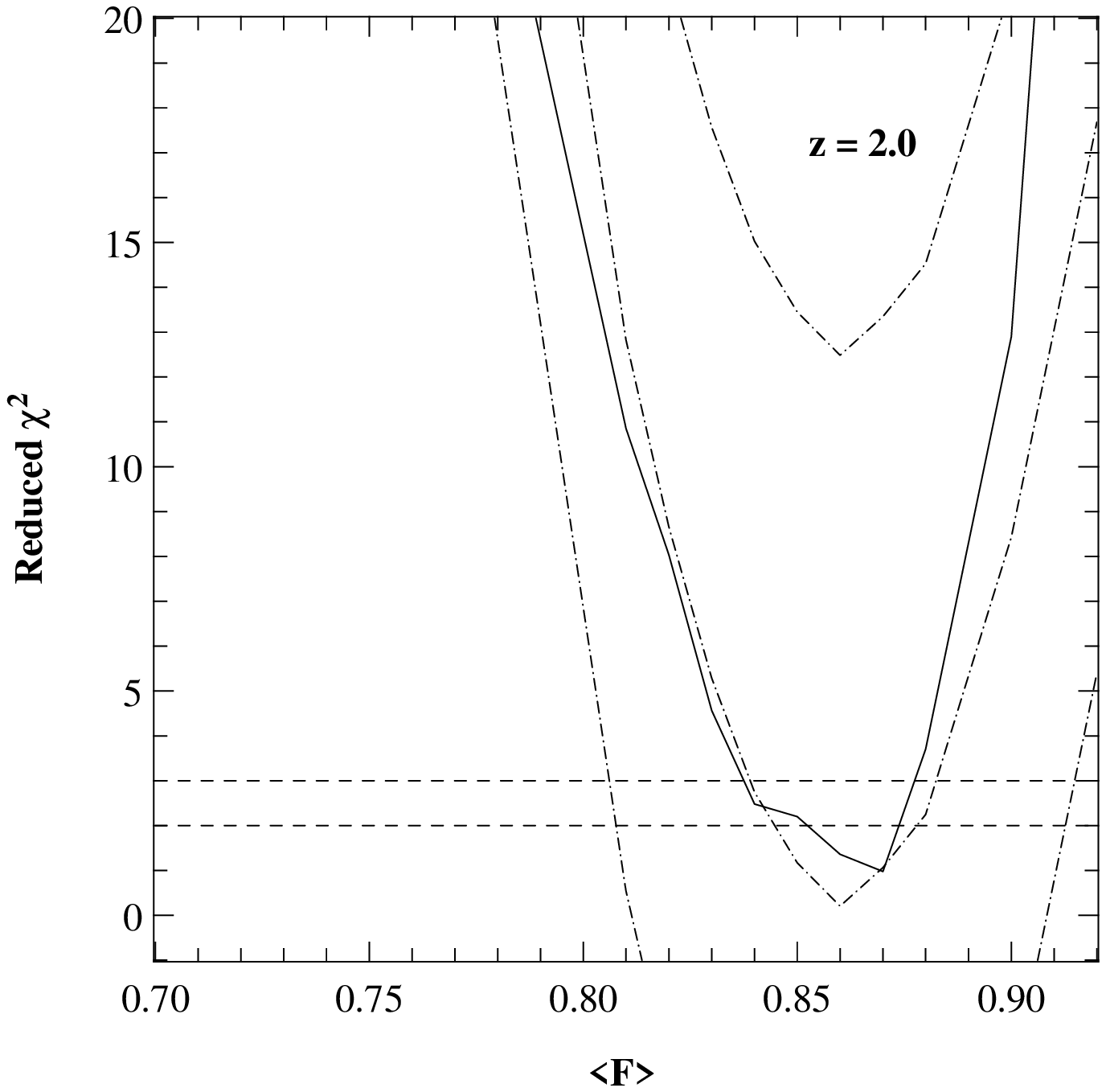}}
\end{picture}
\caption{Reduced $\chi^2$ as a function of the ensemble averaged
 $\langle F\rangle$
at $z=2, 2.5, 3.0$ {\bf (top to bottom)}. The covariance matrix
is measured using the variance among \gimic\ mock samples. 
$\chi^2$ corresponds to the difference between one TPDF
 and  the averaged TPDF from 400 \gimic\ mock samples  
assuming different  $\langle F\rangle$.
As a validity check,  the TPDF measured in  one \gimic\ mock sample
with $\langle F\rangle=0.86, 0.77$ and 0.71 at  $z=2, 2.5, 3.0$ 
respectively, is best fitted with the same value for $\langle F\rangle$  
 (dotted lines show the average reduced $\chi^2$ and the
 1$\sigma$ range  among 400 samples).
The evolution of the reduced $\chi^2$ as a function of $\langle F\rangle$
is similar in the case of the observed LP (solid lines).
 }
\label{fig:pdf_chi2}

\end{figure}

\section{Constraints on the mean transmission and the intensity of the ionising background}
\label{s:gamma}
\label{s:discussion}

\begin{figure}
  \includegraphics[width=\columnwidth]{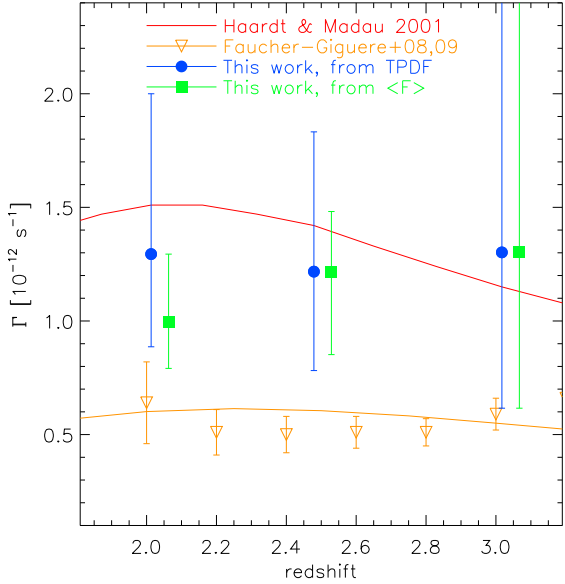}
  \caption{Mean hydrogen photo-ionization rate, $\Gamma$, as function
    of redshift, from summing over sources as computed by Haardt \&
    Madau (2001, {\em red}) and Faucher-Gigu\`ere et al (2009, {\em
      drawn orange line}), and from comparing simulated to observed
    mock spectra. {\em Blue} and {\em green} points are our (2$\sigma$)
    determinations from comparing, respectively, the TPDF and the mean
    flux in the \gimic\ simulations to the LP data, {\em orange}
    symbols are the Faucher-Gigu\`ere et al. (2008) determination using
    a sample of 84 high-resolution quasars.}
\label{fig:gamma}
\end{figure}
\begin{table}
  \caption{{\em Upper rows}: Measured value of the mean transmission  in three redshift ranges for  LUQAS and LP samples.  Using the LP as a reference,  the redshifts ranges are $1.88-2.37$, $2.37-2.71$, $2.71-3.21$  with absorption distance of 10.3, 5.8 and 2.9 respectively.  For LUQAS, Kim et al. (2007) provide errors  computed by bootstrapping chunks  of size 5\AA\ within bins of size $dz=0.2$ ; their errors are then rescaled to the LP absorption distances. The errors given for LP correspond to the variance between \gimic\ mock samples. {\em Lower rows}: ensemble-averaged $\langle F\rangle$ in \gimic\ simulations that reproduce within  2$\sigma$ the LP observed  transmission PDF and  mean transmission, $\bar F$;  the last row gives the  ionising background rate values in the same \gimic\ simulations. $\langle F\rangle$ refers to an ensemble average, $\bar F$ refers to a single realisation of such an ensemble, and is generally larger than $\langle F\rangle$ because it includes a 2\% continuum fitting offset.}
\begin{center}
\begin{tabular}{|c|c|c|r|}
\hline
$\bar z=2.0$ &  $\bar z=2.5$ & $\bar z=3.0$ &   \\
\hline
\hline
\multicolumn{4}{|c|}{Measured $\bar F$ ($\pm 2\sigma$)}\\
\hline
0.887$\pm$ 0.011 & 0.812$\pm$0.017 & 0.780$\pm 0.034$ & LP\\
0.868$\pm$ 0.010 &  0.775$\pm$0.021 & 0.713$\pm$0.032   & LUQAS\\
\hline
\hline
\multicolumn{4}{|c|}{Derived $\langle F\rangle$ with $2\sigma$ variance from \gimic\ mocks.}\\
\hline
0.86$^{+0.007}_{-0.025}$ & 0.77$^{+0.005}_{-0.045}$ & 0.71$^{+0.06}_{-0.05}$ & (from TPDF)  \\
0.85$\pm{0.02}$ & 0.79$^{+0.02}_{-0.04}$ & 0.71$^{+0.07}_{-0.09}$ & (from $\bar F)$ \\
\hline
\hline
\multicolumn{4}{|c|}{Derived $\Gamma_{12}$ with $2\sigma$ range from \gimic\ mocks.}\\
\hline
1.3 (0.9,~2.0) & 1.2 (0.8,~1.8)& 1.3 (0.6,~2.6)  &  \\
\hline
\end{tabular}
\end{center}
\label{tab:meanf}
\end{table}

The photo-ionization rate can be estimated by scaling mock spectra
obtained from simulations to the observed mean transmission $\bar F$,
and calculating the corresponding value of $\Gamma_{12}$. To determine
the range of $\Gamma_{12}$ values consistent with the observed $\bar F$,
we need some measure of the expected variance of $\bar F$ around its
ensemble average $\langle F\rangle$. In principle, it should also be
possible to use the full transmission PDF rather than just its mean.

To judge how well a given realisation of a mock transmission PDF fits
an observational determination, one could use the usual
$\chi^2$-estimator for values of the transmission between 0.1 and
0.7.
A covariance matrix can be
computed by cross-correlating estimates of the TPDF from a large number
of bootstrap samples, as described in \cite{Lidzal06}. Note that all
bootstrap samples are then by construction sub-samples of the observed
spectra, which limits their usefulness if the observed path length is
small. When this is applied to the transmission PDF, it transpires that
the covariance matrix is nearly singular and hence needs to be \lq
regularized\rq\ using a singular value decomposition. We found that the
values obtained for $\chi^2$ then depend strongly on the number of
singular values regularised, which severely compromises the usual
statistical interpretation of $\chi^2$. We can get around this problem
by using the simulations to estimate the variance on either $\bar F$ or
the transmission PDF, for samples with given $\langle F\rangle$.

However, we have seen that the value of the mean transmission $\bar F$
for a given realisation of a mock sample can differ considerably from
the ensemble average $\langle F\rangle$ of the sample. Since the
observations only provide a single measurement of $\bar F$, a
potentially large range of ensemble averages are consistent with that
$\bar F$. This is illustrated in Fig.~\ref{fig:pdf_favg} for the
transmission PDF, and in Fig.~\ref{fig:fW_favg} for $\bar F$, both at
redshift $z=2.5$. In both cases the dark grey band shows the 2$\sigma$
range  in mock samples drawn from simulations with a given
value of the ensemble averaged transmission ($\langle
F\rangle=0.77$ and 0.79 respectively). As before, each sample has the same redshift path as
the LP sample.

Considering first the mean transmission as a function of line-width, 
  we demand   the mean transmission with $W=\infty$ 
 to fall within the 2$\sigma$ region. We interpret
these extreme values as 2$~\sigma$ limits on the ensemble average
$\langle F\rangle$. The 2~$\sigma$ allowed range is then
 $0.75\le\langle F\rangle\le 0.81$. As before, the
determination of $\bar F$ in the mock sample is done after \lq
continuum fitting\rq, which implies that $\bar F$ will be
systematically higher than $\langle F \rangle$.
Performing the same analysis at $z=3$ and at $z=2.$
 yields a 2~$\sigma$ allowed range
of $0.62\le\langle F\rangle\le 0.78$ and $0.83\le\langle F\rangle\le 0.87$ 
respectively (Table~\ref{tab:meanf}).

To do a fit of the TPDF requires a measure of  the covariance matrix. As 
explained  above,  data samples are not yet  large enough to provide a reliable estimate of it. 
Rather, we compute the covariance using 400 independent determinations of the TPDF
in \gimic\ mock samples. The covariance matrix can thus  
be inverted without further regularization. 
We use  13 bins for a range of flux $0.1<F<0.7$, corresponding 
to $k=12$ degree of freedoms. The evolution of the reduced 
$\chi^2_r=(\chi^2-k)/\sqrt(2k)$ is shown in Fig.~\ref{fig:pdf_chi2} (solid lines). To check the validity of this procedure, we derive the same evolution for different mock samples.
Assuming a true value of $\langle F\rangle_{\rm true}$ (0.71, 0.77 and 0.86 at $z=$2, 2.5 and 3 respectively), we compare again 400 mock samples 
with different value of  $\langle F\rangle$ to the average TPDF with
 $\langle F\rangle_{\rm true}$, and compute the associated
 reduced $\chi^2_r$. The average evolution of $\chi^2_r$ and its dispersion
(dotted lines in Fig.~\ref{fig:pdf_chi2}) are consistent with the observed evolution
using the LP TPDF, despite a slight tension at $z=2.5$.
We provide a best fitting value and  a 2$\sigma$ range for $\langle F\rangle$ 
using the smooth average  evolution in \gimic\ samples:
 $0.845\le\langle F\rangle=0.86\le 0.877$ at $z=2.0$,  $0.745\le\langle F\rangle=0.77\le 0.795$ at $z=2.5$ and  $0.66\le\langle F\rangle=0.71\le 0.77$ at $z=3.0$. Note that the best fitting value for $\langle F\rangle$ is slightly shifted compared to the value corresponding to the observed minimum, in order to best reproduce the overall evolution of $\chi^2_r$. Also, the range at $z=2.5$ as determined from the 
evolution of $\chi^2$ is narrower than the range determined by eye in Fig.~\ref{fig:pdf_favg}.
Those 
estimates for $\langle F\rangle$ and their $2~\sigma$ uncertainty at
these three redshifts can be compared
to the values given in
Table~\ref{tab:meanf} that refer to the allowed range 
of $\langle F\rangle$ so that  GIMIC simulations
reproduce within $2\sigma$ the LP observed transmission PDF 
 (Fig.~\ref{fig:pdf_favg}).
 Our values are generally in agreement with
previously published values, but our quoted uncertainties are
significantly larger.  

Given the constraints on $\langle F\rangle$, we can use the simulations
to infer the corresponding range in photo-ionization rates $\Gamma(z)$,
which, in addition to the inferred value of $\langle F\rangle$, depend
on the baryon density, $\Omega_b$, the temperature-density relation,
the fluctuation amplitude $\sigma_8$ and other cosmological parameters
\citep{Rauch97}.

Our inferred values for the photo-ionization rate, $\Gamma(z)$, are
compared in Fig.~\ref{fig:gamma} to the results of Haardt \& Madau
(2001) and to those of \cite{FG2008,FG2009}, and are also listed in
Table~\ref{tab:meanf}. The red \citep{HM01} and orange \citep{FG2009}
curves combine observationally inferred values for the emissivities of
sources of ionising photons with 
an assumed escape fraction and a model for the mean free path based
on observations to estimate $\Gamma$.  Note that \cite{Haardtal11} 
derived recently a lower value of  $\Gamma\simeq 0.9\,10^{-12}{\rm s}^{-1}$ 
for $2<z<3$.
In agreement with these models,
we find little evidence for evolution in $\Gamma$ over the redshift
range $z=$2--3. This is also in agreement with the results
of \citet[][\ , their Figure 7]{Bolton05}, although our error bars are 
again larger for $z=2.5$ and 3.  
Our value for the amplitude is in good agreement with
that from Haardt \& Madau, but is a factor of $\sim 2$ larger
than that of Faucher-Gigu\`ere et al. (2009). The latter value is not
inferred from simulations, but from a fit to the density distribution
of the IGM by \cite{Miralda00}, itself guided by older simulations of
\cite{Miralda1996}. The significant differences in cosmological
parameters of those simulations might explain the significant offset in
the inferred amplitude. Indeed, \cite{pawlik09} found that the
Miralda-Escud\'e et al. fit did not describe their own  simulations well.

\section{Discussion and conclusions}
We have compared the mean transmission, $\bar F$, as well as the
transmission probability distribution function, TPDF, in the \HI\ \lya\ forest
 as derived from several observational samples, as well as from
mock samples computed using the \gimic\ suite of hydrodynamical
simulations. The mean transmission $\bar F$ in the \lya\ forest varies
considerably from QSO to QSO, even at a given redshift. We have shown
that, both in data and in simulations, this is due to the presence of
strong lines, which, though relatively rare, contribute significantly
to the opacity. This implies that a large redshift path is required to
accurately determine the mean transmission.

 We have compared in detail the variance $\sigma$ on $\bar F$
  between published data, our own analysis of the observed UVES LP sample, and
  mocks computed from the \gimic\ hydrodynamical simulations.  We have
  shown, {\em from observations only,} that bootstrap errors depend
  sensitively on chunk size, and only start to converge when relatively
  large chunks, $\gtsima 25$~\AA, are used. This is larger than
  typically used, and as a consequence we claim that published errors
  may be slightly underestimated, especially at larger redshift.
We compared the mean transmission computed from the \gimic\ simulations
to that obtained from three observational samples. The \gimic\
simulations are zoomed simulations of different density regions picked
from the Millennium simulation, and as such they have a realistic
amount of \lq sample variance\rq. We exploited this feature of the
simulations to estimate the uncertainty in the determination
of $\langle F\rangle$ for various observed samples. 
 When
  we compute errors in the same way as performed in published work, we
  find excellent agreement between published and predicted values.
 We have also
  shown that converged bootstrap errors are in good agreement with
  errors found from bootstrapping {\em mock samples}. Thus, we find
  larger uncertainties than in previous works. For a given value
of $\langle F\rangle$, the variance on the mean transmission is large
enough to make all previously published values consistent within the
scatter.

Using mock
spectra derived from \gimic, we have investigated the dependence of the
variance of the mean transmitted flux on the absorption path $\Delta
X$, see Table~\ref{tab:var_length}. At $z=2.5$, with a sample twice as
large as the LP sample, the 2$\sigma$ variance is only 0.013 and
decreases down to 0.009 with a sample four times as large,  which
  is half of the value for 2$\sigma$ for one LP sample, as
  expected. We note, however, that the size of our simulations may not
be sufficient to evaluate the variance with such a large velocity path,
especially at $z=2$.

We have also investigated the probability distribution of the
transmission.
 The ensemble variance between mock samples is 
systematically larger than
the jack-knife errors used by previous authors, by
 a factor of 1.5--2 in the redshift bins ${\bar{z}}=3$. 
More importantly, the covariance matrix derived from a suite of mocks
can be inverted without regularization, contrary to standard estimate
with jack-knife methods.
We used these larger errors 
 and compare data to simulations.

%, we have
%examined whether an \lq inverted\rq\ temperature-density relation is
%required by the data.
 The temperature-density relation,
$T=T_0\,(\rho/\langle \rho\rangle)^{\gamma-1}$, in the \gimic\
simulations is a result of adiabatic cooling and photo-heating due to
an imposed ionising-background as computed by \cite{HM01}, tweaked to
yield values for $T_0$ and $\gamma$ consistent with the measured values
of \cite{Schaye00}. In this model $\gamma>1$ at all times, with a
minimum value of $\gamma\simeq 1.3$ around redshift $z=3$ caused by
HeII re-ionization \citep{Theuns02a}. The \gimic\ transmission PDF is in
agreement with that measured from high-resolution quasar spectra over
the redshift range $z=$2--3 in the transmission range $0.1<F<0.7$. For
$F < 0.1$ there may be differences due to the neglect of
self-shielding in the simulations, whereas for $F> 0.7$ uncertainties
in continuum fitting the data complicate the comparison. 
 This agreement is obtained using a specific set of cosmological
parameters. In particular, we assume  $\sigma_8=0.9$.
The goal of this work is not to provide the best fitting cosmological model,
but to point out the large effect of sample variance. 
Indeed, 
 our model with $(\sigma_8,\gamma)=(0.9,1)$
is not ruled out by the current set of data, while 
\cite{Viel09} discard those values 
at more than 2$\sigma$ when considering the whole flux range.
Thus, we argue that previous suggestions for
  an inverted T-$\rho$ relation may have  resulted from an
  underestimate of the errors in the observations, rather than a discrepancy between data and  the  standard model.

\begin{table}
  \caption{Dependence of the variance of the mean transmission on absorption distance $\Delta X$, for three redshifts. The top row shows the variance (2$\sigma$) for the current LP sample (with given absorption distance $\Delta X$ LP). The second and third rows are for samples two, and four times as large. Errors correspond to the variance among mock LP samples. }
\begin{center}
\begin{tabular}{|c||c|c|c|c|}
\hline
                         & $\bar z=2.0$ &  $\bar z=2.5$ & $\bar z=3.0$ &  \\
 $\Delta X$  LP & 10.5               &  5.8                & 2.9                 & sample size \\
\hline
\hline
 & 0.011 & 0.017 & 0.034 & $\Delta X\times 1$\\ 
& 0.0078 & 0.013 & 0.024 & $\Delta X\times 2$\\ 
& 0.0054 & 0.0088 & 0.017 & $\Delta X\times 4$\\
\hline
\end{tabular}
\end{center}
\label{tab:var_length}
\end{table}

\section*{Acknowledgements}  We thank the anonymous referee for
  useful comments that improved the quality of the paper. We would
like to thank our collaborators to allow us to analyse the \gimic\
simulations for this purpose.  These simulations were carried out using
the HPCx facility at the Edinburgh Parallel Computing Centre (EPCC) as
part of the EC's DEISA \lq Extreme Computing Initiative\rq, and with
the Cosmology Machine at the Institute for Computational Cosmology of
Durham University.  This work was supported by an NWO VIDI grant and by
the Marie Curie Initial training Network CosmoComp
(PITN-GA-2009-238536).

%%% 
\nocite{Wiersma09a}
\nocite{Wiersma09b}

\bibliographystyle{mn2e}
\bibliography{RTPal}

\begin{thebibliography}{}

\bibitem[\protect\citeauthoryear{{Altay}, {Theuns}, {Schaye}, {Crighton} \&
  {Dalla Vecchia}}{{Altay} et~al.}{2011}]{Altay2011}
{Altay} G.,  {Theuns} T.,  {Schaye} J.,  {Crighton} N.~H.~M.,    {Dalla
  Vecchia} C.,  2011, \apjl, 737, L37

\bibitem[\protect\citeauthoryear{{Aracil}, {Petitjean}, {Pichon} \&
  {Bergeron}}{{Aracil} et~al.}{2004}]{aracilal04}
{Aracil} B.,  {Petitjean} P.,  {Pichon} C.,    {Bergeron} J.,  2004, \aap, 419,
  811

\bibitem[\protect\citeauthoryear{{Becker}, {Bolton}, {Haehnelt} \&
  {Sargent}}{{Becker} et~al.}{2011}]{Beckeral11}
{Becker} G.~D.,  {Bolton} J.~S.,  {Haehnelt} M.~G.,    {Sargent} W.~L.~W.,
  2011, \mnras, 410, 1096

\bibitem[\protect\citeauthoryear{{Becker}, {Rauch} \& {Sargent}}{{Becker}
  et~al.}{2007}]{Becker07}
{Becker} G.~D.,  {Rauch} M.,    {Sargent} W.~L.~W.,  2007, \apj, 662, 72

\bibitem[\protect\citeauthoryear{{Bergeron}, {Petitjean} \& {Aracil}
  B.}{{Bergeron} et~al.}{2004}]{Bergeron04}
{Bergeron} J.,  {Petitjean} P.,    {Aracil} B. {\em et al.}.,  2004, The
  Messenger, 118, 40

\bibitem[\protect\citeauthoryear{{Bi}, {Boerner} \& {Chu}}{{Bi}
  et~al.}{1992}]{Bi92}
{Bi} H.~G.,  {Boerner} G.,    {Chu} Y.,  1992, \aap, 266, 1

\bibitem[\protect\citeauthoryear{{Bolton}, {Haehnelt}, {Viel} \&
  {Springel}}{{Bolton} et~al.}{2005}]{Bolton05}
{Bolton} J.~S.,  {Haehnelt} M.~G.,  {Viel} M.,    {Springel} V.,  2005, \mnras,
  357, 1178

\bibitem[\protect\citeauthoryear{{Bolton}, {Oh} \& {Furlanetto}}{{Bolton}
  et~al.}{2009}]{Boltonal09}
{Bolton} J.~S.,  {Oh} S.~P.,    {Furlanetto} S.~R.,  2009, \mnras, 395, 736

\bibitem[\protect\citeauthoryear{{Bolton}, {Viel}, {Kim}, {Haehnelt} \&
  {Carswell}}{{Bolton} et~al.}{2008}]{Bolton08}
{Bolton} J.~S.,  {Viel} M.,  {Kim} T.-S.,  {Haehnelt} M.~G.,    {Carswell}
  R.~F.,  2008, \mnras, 386, 1131

\bibitem[\protect\citeauthoryear{{Boyarsky}, {Ruchayskiy} \&
  {Iakubovskyi}}{{Boyarsky} et~al.}{2009}]{boyarskyal09}
{Boyarsky} A.,  {Ruchayskiy} O.,    {Iakubovskyi} D.,  2009, Journal of
  Cosmology and Astro-Particle Physics, 3, 5

\bibitem[\protect\citeauthoryear{{Calura}, {Tescari}, {D'Odorico}, {Viel},
  {Cristiani}, {Kim} \& {Bolton}}{{Calura} et~al.}{2012}]{Caluraal12}
{Calura} F.,  {Tescari} E.,  {D'Odorico} V.,  {Viel} M.,  {Cristiani} S.,
  {Kim} T.-S.,    {Bolton} J.~S.,  2012, \mnras, 422, 3019

\bibitem[\protect\citeauthoryear{{Carswell}, {Webb}, {Baldwin} \&
  {Atwood}}{{Carswell} et~al.}{1987}]{Carswell87}
{Carswell} R.~F.,  {Webb} J.~K.,  {Baldwin} J.~A.,    {Atwood} B.,  1987, \apj,
  319, 709

\bibitem[\protect\citeauthoryear{{Cen}, {Miralda-Escud{\'e}}, {Ostriker} \&
  {Rauch}}{{Cen} et~al.}{1994}]{Cen94}
{Cen} R.,  {Miralda-Escud{\'e}} J.,  {Ostriker} J.~P.,    {Rauch} M.,  1994,
  \apjl, 437, L9

\bibitem[\protect\citeauthoryear{{Chang}, {Broderick} \& {Pfrommer}}{{Chang}
  et~al.}{2011}]{chang2011}
{Chang} P.,  {Broderick} A.~E.,    {Pfrommer} C.,  2011,
  ArXiv:astro-ph/1106.5504

\bibitem[\protect\citeauthoryear{{Cowie}, {Songaila}, {Kim} \& {Hu}}{{Cowie}
  et~al.}{1995}]{Cowie95}
{Cowie} L.~L.,  {Songaila} A.,  {Kim} T.-S.,    {Hu} E.~M.,  1995, \aj, 109,
  1522

\bibitem[\protect\citeauthoryear{{Crain}, {Theuns} \& {Dalla Vecchia}
  C.}{{Crain} et~al.}{2009}]{Crain09}
{Crain} R.~A.,  {Theuns} T.,    {Dalla Vecchia} C. {\em et al.}.,  2009,
  \mnras, 399, 1773

\bibitem[\protect\citeauthoryear{{Crighton}, {Morris}, {Bechtold}, {Crain},
  {Jannuzi}, {Shone} \& {Theuns}}{{Crighton} et~al.}{2010}]{Crighton2010}
{Crighton} N.~H.~M.,  {Morris} S.~L.,  {Bechtold} J.,  {Crain} R.~A.,
  {Jannuzi} B.~T.,  {Shone} A.,    {Theuns} T.,  2010, \mnras, 402, 1273

\bibitem[\protect\citeauthoryear{{Dalla Vecchia} \& {Schaye}}{{Dalla Vecchia}
  \& {Schaye}}{2008}]{DallaVecchiaandSchaye}
{Dalla Vecchia} C.,  {Schaye} J.,  2008, \mnras, 387, 1431

\bibitem[\protect\citeauthoryear{{Desjacques}, {Nusser} \&
  {Sheth}}{{Desjacques} et~al.}{2007}]{Desjacques2007}
{Desjacques} V.,  {Nusser} A.,    {Sheth} R.~K.,  2007, \mnras, 374, 206

\bibitem[\protect\citeauthoryear{{Fan}, {Carilli} \& {Keating}}{{Fan}
  et~al.}{2006}]{Fan06}
{Fan} X.,  {Carilli} C.~L.,    {Keating} B.,  2006, \araa, 44, 415

\bibitem[\protect\citeauthoryear{{Faucher-Gigu{\`e}re}, {Lidz}, {Zaldarriaga}
  \& {Hernquist}}{{Faucher-Gigu{\`e}re} et~al.}{2009}]{FG2009}
{Faucher-Gigu{\`e}re} C.,  {Lidz} A.,  {Zaldarriaga} M.,    {Hernquist} L.,
  2009, \apj, 703, 1416

\bibitem[\protect\citeauthoryear{{Faucher-Gigu{\`e}re}, {Prochaska}, {Lidz},
  {Hernquist} \& {Zaldarriaga}}{{Faucher-Gigu{\`e}re} et~al.}{2008}]{FG2008}
{Faucher-Gigu{\`e}re} C.-A.,  {Prochaska} J.~X.,  {Lidz} A.,  {Hernquist} L.,
   {Zaldarriaga} M.,  2008, \apj, 681, 831

\bibitem[\protect\citeauthoryear{{Fukugita}, {Hogan} \& {Peebles}}{{Fukugita}
  et~al.}{1998}]{Fukugita}
{Fukugita} M.,  {Hogan} C.~J.,    {Peebles} P.~J.~E.,  1998, \apj, 503, 518

\bibitem[\protect\citeauthoryear{{Gratton}, {Lewis} \& {Efstathiou}}{{Gratton}
  et~al.}{2008}]{grattonal08}
{Gratton} S.,  {Lewis} A.,    {Efstathiou} G.,  2008, Physical Review D, 77,
  083507

\bibitem[\protect\citeauthoryear{{Guimar{\~a}es}, {Petitjean}, {Rollinde}, {de
  Carvalho}, {Djorgovski}, {Srianand}, {Aghaee} \& {Castro}}{{Guimar{\~a}es}
  et~al.}{2007}]{Guimaraes}
{Guimar{\~a}es} R.,  {Petitjean} P.,  {Rollinde} E.,  {de Carvalho} R.~R.,
  {Djorgovski} S.~G.,  {Srianand} R.,  {Aghaee} A.,    {Castro} S.,  2007,
  \mnras, 377, 657

\bibitem[\protect\citeauthoryear{{Gunn} \& {Peterson}}{{Gunn} \&
  {Peterson}}{1965}]{GunnandPeterson}
{Gunn} J.~E.,  {Peterson} B.~A.,  1965, \apj, 142, 1633

\bibitem[\protect\citeauthoryear{{Haardt} \& {Madau}}{{Haardt} \&
  {Madau}}{2001}]{HM01}
{Haardt} F.,  {Madau} P.,  2001, in {Neumann} D.~M.,  {Tran} J.~T.~V.,  eds,
  Clusters of Galaxies and the High Redshift Universe Observed in X-rays
  {Modelling the UV/X-ray cosmic background with CUBA}

\bibitem[\protect\citeauthoryear{{Haardt} \& {Madau}}{{Haardt} \&
  {Madau}}{2011}]{Haardtal11}
{Haardt} F.,  {Madau} P.,  2011, ArXiv:1105.2039

\bibitem[\protect\citeauthoryear{{Hernquist}, {Katz}, {Weinberg} \&
  {Miralda-Escud{\'e}}}{{Hernquist} et~al.}{1996}]{Hernquist96}
{Hernquist} L.,  {Katz} N.,  {Weinberg} D.~H.,    {Miralda-Escud{\'e}} J.,
  1996, \apjl, 457, L51

\bibitem[\protect\citeauthoryear{{Hu}, {Kim}, {Cowie}, {Songaila} \&
  {Rauch}}{{Hu} et~al.}{1995}]{Hu95}
{Hu} E.~M.,  {Kim} T.-S.,  {Cowie} L.~L.,  {Songaila} A.,    {Rauch} M.,  1995,
  \aj, 110, 1526

\bibitem[\protect\citeauthoryear{{Hui} \& {Gnedin}}{{Hui} \&
  {Gnedin}}{1997}]{HuiandGnedin}
{Hui} L.,  {Gnedin} N.~Y.,  1997, \mnras, 292, 27

\bibitem[\protect\citeauthoryear{{Hui} \& {Haiman}}{{Hui} \&
  {Haiman}}{2003}]{Hui03}
{Hui} L.,  {Haiman} Z.,  2003, \apj, 596, 9

\bibitem[\protect\citeauthoryear{{Kim}, {Bolton}, {Viel}, {Haehnelt} \&
  {Carswell}}{{Kim} et~al.}{2007}]{Kim07}
{Kim} T.-S.,  {Bolton} J.~S.,  {Viel} M.,  {Haehnelt} M.~G.,    {Carswell}
  R.~F.,  2007, \mnras, 382, 1657

\bibitem[\protect\citeauthoryear{{Kim} \& {Croft}}{{Kim} \&
  {Croft}}{2008}]{KimCroft}
{Kim} Y.-R.,  {Croft} R.~A.~C.,  2008, \mnras, 387, 377

\bibitem[\protect\citeauthoryear{{Kirkman}, {Tytler}, {Suzuki}, {Melis},
  {Hollywood}, {James}, {So}, {Lubin}, {Jena}, {Norman} \& {Paschos}}{{Kirkman}
  et~al.}{2005}]{Kirkman2005}
{Kirkman} D.,  {Tytler} D.,  {Suzuki} N.,  {Melis} C.,  {Hollywood} S.,
  {James} K.,  {So} G.,  {Lubin} D.,  {Jena} T.,  {Norman} M.~L.,    {Paschos}
  P.,  2005, \mnras, 360, 1373

\bibitem[\protect\citeauthoryear{{Komatsu}, {Dunkley}, {Nolta}, {Bennett},
  {Gold}, {Hinshaw}, {Jarosik}, {Larson}, {Limon}, {Page}, {Spergel},
  {Halpern}, {Hill}, {Kogut}, {Meyer}, {Tucker}, {Weiland}, {Wollack} \&
  {Wright}}{{Komatsu} et~al.}{2009}]{Komatsual09}
{Komatsu} E.,  {Dunkley} J.,  {Nolta} M.~R.,  {Bennett} C.~L.,  {Gold} B.,
  {Hinshaw} G.,  {Jarosik} N.,  {Larson} D.,  {Limon} M.,  {Page} L.,
  {Spergel} D.~N.,  {Halpern} M.,  {Hill} R.~S.,  {Kogut} A.,  {Meyer} S.~S.,
  {Tucker} G.~S.,  {Weiland} J.~L.,  {Wollack} E.,    {Wright} E.~L.,  2009,
  \apjs, 180, 330

\bibitem[\protect\citeauthoryear{{Lidz}, {Faucher-Gigu{\`e}re}, {Dall'Aglio},
  {McQuinn}, {Fechner}, {Zaldarriaga}, {Hernquist} \& {Dutta}}{{Lidz}
  et~al.}{2010}]{Lidzal10}
{Lidz} A.,  {Faucher-Gigu{\`e}re} C.-A.,  {Dall'Aglio} A.,  {McQuinn} M.,
  {Fechner} C.,  {Zaldarriaga} M.,  {Hernquist} L.,    {Dutta} S.,  2010, \apj,
  718, 199

\bibitem[\protect\citeauthoryear{{Lidz}, {Heitmann}, {Hui}, {Habib}, {Rauch} \&
  {Sargent}}{{Lidz} et~al.}{2006}]{Lidzal06}
{Lidz} A.,  {Heitmann} K.,  {Hui} L.,  {Habib} S.,  {Rauch} M.,    {Sargent}
  W.~L.~W.,  2006, \apj, 638, 27

\bibitem[\protect\citeauthoryear{{Lynds}}{{Lynds}}{1971}]{Lynds}
{Lynds} R.,  1971, \apjl, 164, L73

\bibitem[\protect\citeauthoryear{{McDonald} \& {Miralda-Escud{\'e}}}{{McDonald}
  \& {Miralda-Escud{\'e}}}{1999}]{McDonald99}
{McDonald} P.,  {Miralda-Escud{\'e}} J.,  1999, \apj, 518, 24

\bibitem[\protect\citeauthoryear{{McDonald}, {Miralda-Escud{\'e}} \& {Rauch}
  M.}{{McDonald} et~al.}{2000}]{McDonald00}
{McDonald} P.,  {Miralda-Escud{\'e}} J.,    {Rauch} M. {\em et al.}.,  2000,
  \apj, 543, 1

\bibitem[\protect\citeauthoryear{{McDonald}, {Miralda-Escud{\'e}}, {Rauch},
  {Sargent}, {Barlow} \& {Cen}}{{McDonald} et~al.}{2001}]{McDonald01}
{McDonald} P.,  {Miralda-Escud{\'e}} J.,  {Rauch} M.,  {Sargent} W.~L.~W.,
  {Barlow} T.~A.,    {Cen} R.,  2001, \apj, 562, 52

\bibitem[\protect\citeauthoryear{{McDonald}, {Seljak} \& {Burles}}{{McDonald}
  et~al.}{2006}]{McDonald06}
{McDonald} P.,  {Seljak} U.,    {Burles} {\em et al.}.,  2006, \apjs, 163, 80

\bibitem[\protect\citeauthoryear{{McDonald}, {Seljak}, {Cen}, {Bode} \&
  {Ostriker}}{{McDonald} et~al.}{2005}]{McDonald05}
{McDonald} P.,  {Seljak} U.,  {Cen} R.,  {Bode} P.,    {Ostriker} J.~P.,  2005,
  \mnras, 360, 1471

\bibitem[\protect\citeauthoryear{{McQuinn}, {Hernquist}, {Lidz} \&
  {Zaldarriaga}}{{McQuinn} et~al.}{2011}]{McQuinnal11}
{McQuinn} M.,  {Hernquist} L.,  {Lidz} A.,    {Zaldarriaga} M.,  2011, \mnras,
  415, 977

\bibitem[\protect\citeauthoryear{{McQuinn}, {Lidz}, {Zaldarriaga}, {Hernquist},
  {Hopkins}, {Dutta} \& {Faucher-Gigu{\`e}re}}{{McQuinn}
  et~al.}{2009}]{McQuinn09}
{McQuinn} M.,  {Lidz} A.,  {Zaldarriaga} M.,  {Hernquist} L.,  {Hopkins} P.~F.,
   {Dutta} S.,    {Faucher-Gigu{\`e}re} C.-A.,  2009, \apj, 694, 842

\bibitem[\protect\citeauthoryear{{Meiksin}, {Bryan} \& {Machacek}}{{Meiksin}
  et~al.}{2001}]{Meiksin2001}
{Meiksin} A.,  {Bryan} G.,    {Machacek} M.,  2001, \mnras, 327, 296

\bibitem[\protect\citeauthoryear{{Miralda-Escud{\'e}}, {Cen}, {Ostriker} \&
  {Rauch}}{{Miralda-Escud{\'e}} et~al.}{1996}]{Miralda1996}
{Miralda-Escud{\'e}} J.,  {Cen} R.,  {Ostriker} J.~P.,    {Rauch} M.,  1996,
  \apj, 471, 582

\bibitem[\protect\citeauthoryear{{Miralda-Escud{\'e}}, {Haehnelt} \&
  {Rees}}{{Miralda-Escud{\'e}} et~al.}{2000}]{Miralda00}
{Miralda-Escud{\'e}} J.,  {Haehnelt} M.,    {Rees} M.~J.,  2000, \apj, 530, 1

\bibitem[\protect\citeauthoryear{{Mortlock}, {Warren}, {Venemans}, {Patel},
  {Hewett}, {McMahon}, {Simpson}, {Theuns}, {Gonz{\'a}les-Solares}, {Adamson},
  {Dye}, {Hambly}, {Hirst}, {Irwin}, {Kuiper}, {Lawrence} \&
  {R{\"o}ttgering}}{{Mortlock} et~al.}{2011}]{mortlock11}
{Mortlock} D.~J.,  {Warren} S.~J.,  {Venemans} B.~P.,  {Patel} M.,  {Hewett}
  P.~C.,  {McMahon} R.~G.,  {Simpson} C.,  {Theuns} T.,  {Gonz{\'a}les-Solares}
  E.~A.,  {Adamson} A.,  {Dye} S.,  {Hambly} N.~C.,  {Hirst} P.,  {Irwin}
  M.~J.,  {Kuiper} E.,  {Lawrence} A.,    {R{\"o}ttgering} H.~J.~A.,  2011,
  \nat, 474, 616

\bibitem[\protect\citeauthoryear{{Pawlik}, {Schaye} \& {van
  Scherpenzeel}}{{Pawlik} et~al.}{2009}]{pawlik09}
{Pawlik} A.~H.,  {Schaye} J.,    {van Scherpenzeel} E.,  2009, \mnras, 394,
  1812

\bibitem[\protect\citeauthoryear{{Petitjean}, {Mueket} \& {Kates}}{{Petitjean}
  et~al.}{1995}]{petitjeanal95}
{Petitjean} P.,  {Mueket} J.~P.,    {Kates} R.~E.,  1995, \aap, 295, L9

\bibitem[\protect\citeauthoryear{{Petitjean}, {Webb}, {Rauch}, {Carswell} \&
  {Lanzetta}}{{Petitjean} et~al.}{1993}]{petitjeanal93}
{Petitjean} P.,  {Webb} J.~K.,  {Rauch} M.,  {Carswell} R.~F.,    {Lanzetta}
  K.,  1993, \mnras, 262, 499

\bibitem[\protect\citeauthoryear{{Puchwein}, {Pfrommer}, {Springel},
  {Broderick} \& {Chang}}{{Puchwein} et~al.}{2011}]{Puchweinal11}
{Puchwein} E.,  {Pfrommer} C.,  {Springel} V.,  {Broderick} A.~E.,    {Chang}
  P.,  2011, ArXiv:astro-ph/1107.3837

\bibitem[\protect\citeauthoryear{{Rauch}}{{Rauch}}{1998}]{Rauch98}
{Rauch} M.,  1998, \araa, 36, 267

\bibitem[\protect\citeauthoryear{{Rauch}, {Miralda-Escude}, {Sargent},
  {Barlow}, {Weinberg}, {Hernquist}, {Katz}, {Cen} \& {Ostriker}}{{Rauch}
  et~al.}{1997}]{Rauch97}
{Rauch} M.,  {Miralda-Escude} J.,  {Sargent} W.~L.~W.,  {Barlow} T.~A.,
  {Weinberg} D.~H.,  {Hernquist} L.,  {Katz} N.,  {Cen} R.,    {Ostriker}
  J.~P.,  1997, \apj, 489, 7

\bibitem[\protect\citeauthoryear{{Ricotti}, {Gnedin} \& {Shull}}{{Ricotti}
  et~al.}{2000}]{Ricotti2000}
{Ricotti} M.,  {Gnedin} N.~Y.,    {Shull} J.~M.,  2000, \apj, 534, 41

\bibitem[\protect\citeauthoryear{{Rollinde}, {Petitjean}, {Pichon}, {Colombi},
  {Aracil}, {D'Odorico} \& {Haehnelt}}{{Rollinde} et~al.}{2003}]{Rollinde03}
{Rollinde} E.,  {Petitjean} P.,  {Pichon} C.,  {Colombi} S.,  {Aracil} B.,
  {D'Odorico} V.,    {Haehnelt} M.~G.,  2003, \mnras, 341, 1279

\bibitem[\protect\citeauthoryear{{Rollinde}, {Srianand}, {Theuns}, {Petitjean}
  \& {Chand}}{{Rollinde} et~al.}{2005}]{Rollinde}
{Rollinde} E.,  {Srianand} R.,  {Theuns} T.,  {Petitjean} P.,    {Chand} H.,
  2005, \mnras, 361, 1015

\bibitem[\protect\citeauthoryear{{Schaye}}{{Schaye}}{2001}]{Schaye2001}
{Schaye} J.,  2001, \apj, 559, 507

\bibitem[\protect\citeauthoryear{{Schaye}, {Aguirre}, {Kim}, {Theuns}, {Rauch}
  \& {Sargent}}{{Schaye} et~al.}{2003}]{Schaye03}
{Schaye} J.,  {Aguirre} A.,  {Kim} T.-S.,  {Theuns} T.,  {Rauch} M.,
  {Sargent} W.~L.~W.,  2003, \apj, 596, 768

\bibitem[\protect\citeauthoryear{{Schaye} \& {Dalla Vecchia}}{{Schaye} \&
  {Dalla Vecchia}}{2008}]{SchayeandDallaVecchia}
{Schaye} J.,  {Dalla Vecchia} C.,  2008, \mnras, 383, 1210

\bibitem[\protect\citeauthoryear{{Schaye}, {Dalla Vecchia}, {Booth}, {Wiersma},
  {Theuns}, {Haas}, {Bertone}, {Duffy}, {McCarthy} \& {van de Voort}}{{Schaye}
  et~al.}{2010}]{schayeal10}
{Schaye} J.,  {Dalla Vecchia} C.,  {Booth} C.~M.,  {Wiersma} R.~P.~C.,
  {Theuns} T.,  {Haas} M.~R.,  {Bertone} S.,  {Duffy} A.~R.,  {McCarthy} I.~G.,
     {van de Voort} F.,  2010, \mnras, 402, 1536

\bibitem[\protect\citeauthoryear{{Schaye}, {Theuns}, {Rauch}, {Efstathiou} \&
  {Sargent}}{{Schaye} et~al.}{2000}]{Schaye00}
{Schaye} J.,  {Theuns} T.,  {Rauch} M.,  {Efstathiou} G.,    {Sargent}
  W.~L.~W.,  2000, \mnras, 318, 817

\bibitem[\protect\citeauthoryear{{Springel}}{{Springel}}{2005}]{Springel05b}
{Springel} V.,  2005, \mnras, 364, 1105

\bibitem[\protect\citeauthoryear{{Springel}, {White}, {Jenkins}, {Frenk},
  {Yoshida}, {Gao}, {Navarro}, {Thacker}, {Croton}, {Helly}, {Peacock}, {Cole},
  {Thomas}, {Couchman}, {Evrard}, {Colberg} \& {Pearce}}{{Springel}
  et~al.}{2005}]{Springel05}
{Springel} V.,  {White} S.~D.~M.,  {Jenkins} A.,  {Frenk} C.~S.,  {Yoshida} N.,
   {Gao} L.,  {Navarro} J.,  {Thacker} R.,  {Croton} D.,  {Helly} J.,
  {Peacock} J.~A.,  {Cole} S.,  {Thomas} P.,  {Couchman} H.,  {Evrard} A.,
  {Colberg} J.,    {Pearce} F.,  2005, \nat, 435, 629

\bibitem[\protect\citeauthoryear{{Theuns}, {Leonard}, {Efstathiou}, {Pearce} \&
  {Thomas}}{{Theuns} et~al.}{1998}]{Theuns98}
{Theuns} T.,  {Leonard} A.,  {Efstathiou} G.,  {Pearce} F.~R.,    {Thomas}
  P.~A.,  1998, \mnras, 301, 478

\bibitem[\protect\citeauthoryear{{Theuns}, {Schaye}, {Zaroubi}, {Kim},
  {Tzanavaris} \& {Carswell}}{{Theuns} et~al.}{2002}]{Theuns02a}
{Theuns} T.,  {Schaye} J.,  {Zaroubi} S.,  {Kim} T.-S.,  {Tzanavaris} P.,
  {Carswell} B.,  2002, \apjl, 567, L103

\bibitem[\protect\citeauthoryear{{Theuns}, {Viel}, {Kay}, {Schaye}, {Carswell}
  \& {Tzanavaris}}{{Theuns} et~al.}{2002}]{Theuns02b}
{Theuns} T.,  {Viel} M.,  {Kay} S.,  {Schaye} J.,  {Carswell} R.~F.,
  {Tzanavaris} P.,  2002, \apjl, 578, L5

\bibitem[\protect\citeauthoryear{{Theuns}, {Zaroubi}, {Kim}, {Tzanavaris} \&
  {Carswell}}{{Theuns} et~al.}{2002}]{Theuns2002c}
{Theuns} T.,  {Zaroubi} S.,  {Kim} T.-S.,  {Tzanavaris} P.,    {Carswell}
  R.~F.,  2002, \mnras, 332, 367

\bibitem[\protect\citeauthoryear{{Tytler}, {Paschos}, {Kirkman}, {Norman} \&
  {Jena}}{{Tytler} et~al.}{2009}]{Tytler2009}
{Tytler} D.,  {Paschos} P.,  {Kirkman} D.,  {Norman} M.~L.,    {Jena} T.,
  2009, \mnras, 393, 723

\bibitem[\protect\citeauthoryear{{Viel}, {Bolton} \& {Haehnelt}}{{Viel}
  et~al.}{2009}]{Viel09}
{Viel} M.,  {Bolton} J.~S.,    {Haehnelt} M.~G.,  2009, \mnras, 399, L39

\bibitem[\protect\citeauthoryear{{Viel} \& {Haehnelt}}{{Viel} \&
  {Haehnelt}}{2006}]{Viel06}
{Viel} M.,  {Haehnelt} M.~G.,  2006, \mnras, 365, 231

\bibitem[\protect\citeauthoryear{{Viel}, {Haehnelt}, {Carswell} \&
  {Kim}}{{Viel} et~al.}{2004}]{Viel04}
{Viel} M.,  {Haehnelt} M.~G.,  {Carswell} R.~F.,    {Kim} T.-S.,  2004, \mnras,
  349, L33

\bibitem[\protect\citeauthoryear{{Wiersma}, {Schaye} \& {Smith}}{{Wiersma}
  et~al.}{2009}]{Wiersma09a}
{Wiersma} R.~P.~C.,  {Schaye} J.,    {Smith} B.~D.,  2009, \mnras, 393, 99

\bibitem[\protect\citeauthoryear{{Wiersma}, {Schaye}, {Theuns}, {Dalla Vecchia}
  \& {Tomatore}}{{Wiersma} et~al.}{2009}]{Wiersma09b}
{Wiersma} R.~P.~C.,  {Schaye} J.,  {Theuns} T.,  {Dalla Vecchia} C.,
  {Tomatore} L.,  2009, \mnras, 399, 574

\bibitem[\protect\citeauthoryear{{Zhang}, {Anninos} \& {Norman}}{{Zhang}
  et~al.}{1995}]{Zhang95}
{Zhang} Y.,  {Anninos} P.,    {Norman} M.~L.,  1995, \apjl, 453, L57

\end{thebibliography}

\end{document}